\documentclass[publish]{aa}  

\usepackage{graphicx}
\usepackage{txfonts}
\usepackage{lipsum}
\usepackage{subcaption}        
\usepackage{hyperref}
\usepackage{rotating}  
\usepackage{ulem}  

\usepackage{pdflscape}             
                                
\usepackage{placeins}

\defcitealias{Kwan2011MNRAS.411.2383K}{KF11}
\usepackage{multirow}

\begin{document}

    \title{Improving accretion diagnostics for young stellar objects with mid-infrared hydrogen lines from JWST/MIRI}

\author{B. Shridharan\inst{1}
    \and P. Manoj\inst{1}
    \and Vinod Chandra Pathak\inst{1}
    \and A. Caratti o Garatti\inst{2}
    \and Bihan Banerjee\inst{1}
    \and Th. Henning\inst{3}
    \and I. Kamp\inst{4}
    \and E. van Dishoeck\inst{5,6}
    \and H. Tyagi\inst{1}
    \and R. Arun\inst{7,8}
    \and B. Mathew\inst{8}
    \and M. Güdel\inst{9,10}
    \and P.-O. Lagage\inst{11}}

\institute{
Department of Astronomy and Astrophysics, Tata Institute of Fundamental Research, Homi Bhabha Road, Mumbai 400005, India\\
\email{shridharan.1997@gmail.com}
\and INAF – Osservatorio Astronomico di Capodimonte, Salita Moiariello 16, 80131 Napoli, Italy
\and Max-Planck-Institut für Astronomie (MPIA), Königstuhl 17, D-69117 Heidelberg, Germany
\and Kapteyn Astronomical Institute, Rijksuniversiteit Groningen, Postbus 800, 9700AV Groningen, The Netherlands
\and Leiden Observatory, Universiteit Leiden, Leiden, Zuid-Holland, The Netherlands
\and Max-Planck-Institut für Extraterrestrische Physik, D-85748 Garching bei München, Germany
\and Indian Institute of Astrophysics, 2nd Block Koramangala, Bangalore 560034, India
\and Department of Physics and Electronics, CHRIST (Deemed to be University), Bangalore 560029, India
\and Department of Astrophysics, University of Vienna, Türkenschanzstr. 17, A-1180 Vienna, Austria
\and ETH Zürich, Institute for Particle Physics and Astrophysics, Wolfgang-Pauli-Str. 27, 8093 Zürich, Switzerland
\and Université Paris-Saclay, Université Paris Cité, CEA, CNRS, AIM, F-91191 Gif-sur-Yvette, France
}

   \date{Received September 30, 20XX}

 \abstract
   {We present a comprehensive study of mid-infrared neutral hydrogen (H~\textsc{i}) emission lines in 79 nearby (d $<$ 200 $pc$) young stars using the \textit{James Webb Space Telescope} (\textit{JWST}) Mid-Infrared Instrument (MIRI). This work extends accretion diagnostics to mid-infrared H~\textsc{i} transitions, which are less affected by extinction and outflow emission compared to optical and near-infrared H~\textsc{i} lines.}
   {We aim to identify mid-infrared H~\textsc{i} transitions that can serve as reliable accretion diagnostics in young stars, and evaluate their utility in deriving physical conditions of the accreting gas.} 
   {We identified and measured 22 H~\textsc{i} transitions in the MIRI wavelength regime (5–28 $\mu m$) and performed LTE slab modelling to remove the H\textsubscript{2}O contribution from selected H~\textsc{i} transitions. We examined the spatial extent of MIR H~\textsc{i} emission and assessed contamination from molecular and jet-related emission.}
   {We find that mid-IR H~\textsc{i} line emission is spatially compact, even for sources with spatially extended [Ne~\textsc{ii}] and [Fe~\textsc{ii}] jets, suggesting minimal contamination from extended jet. Although Pfund~$\alpha$ (H~\textsc{i}~6--5) and Humphreys~$\alpha$ (H~\textsc{i}~7--6) are the strongest lines in the mid-infrared, they are blended with H$_2$O transitions. This blending necessitates additional processing to remove molecular contamination, thereby limiting their use as accretion diagnostics. Instead, we identify the H~\textsc{i}~(8--6) at 7.502 $\mu m$ and H~\textsc{i}~(10--7) at 8.760 $\mu m$ transitions as better alternatives, as they are largely unaffected by molecular contamination and offer a more reliable means of measuring accretion rates from MIRI spectra. We provide updated empirical relations for converting mid-IR H~\textsc{i} line luminosities into accretion luminosity for six different H~\textsc{i} lines in the MIRI wavelength range. Moreover, a comparison of the observed line ratios with theoretical models shows that MIR H~\textsc{i} lines offer robust constraints on the hydrogen gas density in accretion columns, $n_\mathrm{H} = $ 10$^{10.6}$ to 10$^{11.2}$ cm$^{-3}$ in most stars, with some stars exhibiting lower densities ($<10^{10}$~cm$^{-3}$), approaching the optically thin regime.}
   {}


\authorrunning{Shridharan et al. }
\titlerunning{Mid-Infrared Hydrogen Lines as accretion diagnostic from JWST/MIRI}
\maketitle

\section{Introduction}

Accretion in pre-main-sequence (PMS) stars is a fundamental process that significantly influences their evolution and the structure of protoplanetary discs surrounding them. Ultraviolet (UV) excess over the photospheric continuum is considered the primary indicator of accretion, as it traces the hot continuum emission from the accretion shock, where material from the disc falls onto the stellar surface guided by the magnetospheric columns \citep[and others]{Koenigl1991ApJ...370L..39K, Edwards1994AJ....108.1056E,Hartigan1995ApJ...452..736H, muzerolle1998magnetospheric, Calvet1998ApJ...509..802C}. However, due to observational challenges and higher extinction in the UV region, its use as an accretion indicator is typically limited to relatively unobscured PMS stars. Neutral hydrogen (H~\textsc{i}) emission lines, especially the H$\alpha$ (3--2) transition at 0.656 $\mu$m, have been widely used as accretion diagnostics in PMS stars \citep[e.g.][]{Alcala2017AA...600A..20A, Alcala2014AA...561A...2A, Fairlamb2015MNRAS.453..976F, Antoniucci2017AA...599A.105A}. These H~\textsc{i} lines serve as proxies for estimating mass accretion rates through empirical correlations between H~\textsc{i} line luminosities and accretion luminosities derived from UV excess measurements \citep[e.g.][]{Muzerolle1998AJ....116.2965M, Calvet2004AJ....128.1294C,Natta2006AA...452..245N,Herczeg2008ApJ...681..594H,Riglialco2012AA...548A..56R}.

The advent of instruments such as VLT/X-shooter \citep{2011AA...536A.105V} has expanded access to H~\textsc{i} lines across a broad wavelength range, from the near-UV to the near-infrared (NIR), extending their use beyond H$\alpha$ to include NIR lines such as Pa$\beta$, Br$\gamma$, and others \citep{Riglialco2012AA...548A..56R, Ingleby2013ApJ...767..112I, Manara2014AA...568A..18M, Manara2015AA...579A..66M, Fiorellino2023ApJ...944..135F}. The lower opacity and extinction effects in NIR H~\textsc{i} lines compared to optical lines have also enabled the determination of accretion rates for more embedded sources. \citet{Salyk2013ApJ...769...21S}, using NIRSPEC at the Keck II telescope and CRIRES at the Very Large Telescope, introduced the Pf$\beta$ (7--5) line at 4.65 $\mu m$ as an accretion diagnostic. \citet{Rigliaco2015ApJ...801...31R} analysed the H~\textsc{i} (7-6) line at 12.37 $\mu m$ using Spitzer spectra for 118 sources and established a MIR H~\textsc{i} line as an accretion tracer for the first time. More recently, \citet{Tofflemire2025arXiv250408029T} updated the empirical relation for H~\textsc{i} (7–6) and additionally established H~\textsc{i} (10–7) at 8.76~$\mu m$ as a {more reliable accretion indicator, which does not have any significant contamination from molecular lines. }

H~\textsc{i} emission lines from young stars can also be used to derive the physical properties of the circumstellar gas. Previous investigations using VLT/X-shooter observations and other instruments (0.3–2.4~$\mu m$) have systematically studied the Balmer and Paschen series decrements in large, homogeneous samples of T~Tauri stars, which have provided valuable insights into the physical conditions of the emitting gas \citep{Antoniucci2017AA...599A.105A, Edwards2013ApJ...778..148E, Bary2008ApJ...687..376B}. Such analyses have often challenged the applicability of Case~B\footnote{Case
B electron recombination model is applicable where the emitting medium
is assumed to be optically thick to Lyman photons but
is optically thin to photons of all other HI transitions} recombination theory in circumstellar environments, highlighting the need for more comprehensive modelling of the emission regions such as in \citet{Kwan2011MNRAS.411.2383K}. Additionally, comparisons of optical and NIR line profiles indicate a stronger contribution from winds and outflows in optical lines, whereas NIR lines primarily trace the accreting gas \citep{Eisner2015MNRAS.447..202E, Najita1996ApJ...456..292N, Muzerolle2001ApJ...550..944M, Folha2001AA...365...90F}. Extending such analyses to mid-infrared H~\textsc{i} transitions can further mitigate contamination from winds and outflows, offering deeper insights into the physical conditions within the accretion column.

In this work, we compiled \textit{JWST}/MIRI spectra for 79 young stellar objects (YSOs) from the \textit{JWST} data archive to study H~\textsc{i} emission lines in the mid-infrared spectra. Specifically, we analysed 22 H~\textsc{i} transitions with upper level quantum numbers $N_\mathrm{up} < 14$, falling within the MIRI Medium resolution spectrometer (MRS) wavelength range (5--28 $\mu m$). Section~\ref{sec:data} describes the compilation and reduction of \textit{JWST}/MIRI spectra, including the spatial extent of the detected H~\textsc{i} lines. In Section~\ref{sec:analysis}, we detail the detection of H~\textsc{i} lines, removal of H\textsubscript{2}O contamination, and derive updated empirical relations to estimate $L_{\rm acc}$ for six transitions (Section~\ref{sec:empirical}). Here, we also assess the impact of accretion variability and compare with other MIR diagnostics. Section~\ref{sec:physcond} examines the physical conditions of the H~\textsc{i}-emitting gas. Section \ref{sec:summary} provides a summary of the results from our work. 

\begin{figure*}
    \centering
    \includegraphics[width=2.1\columnwidth]{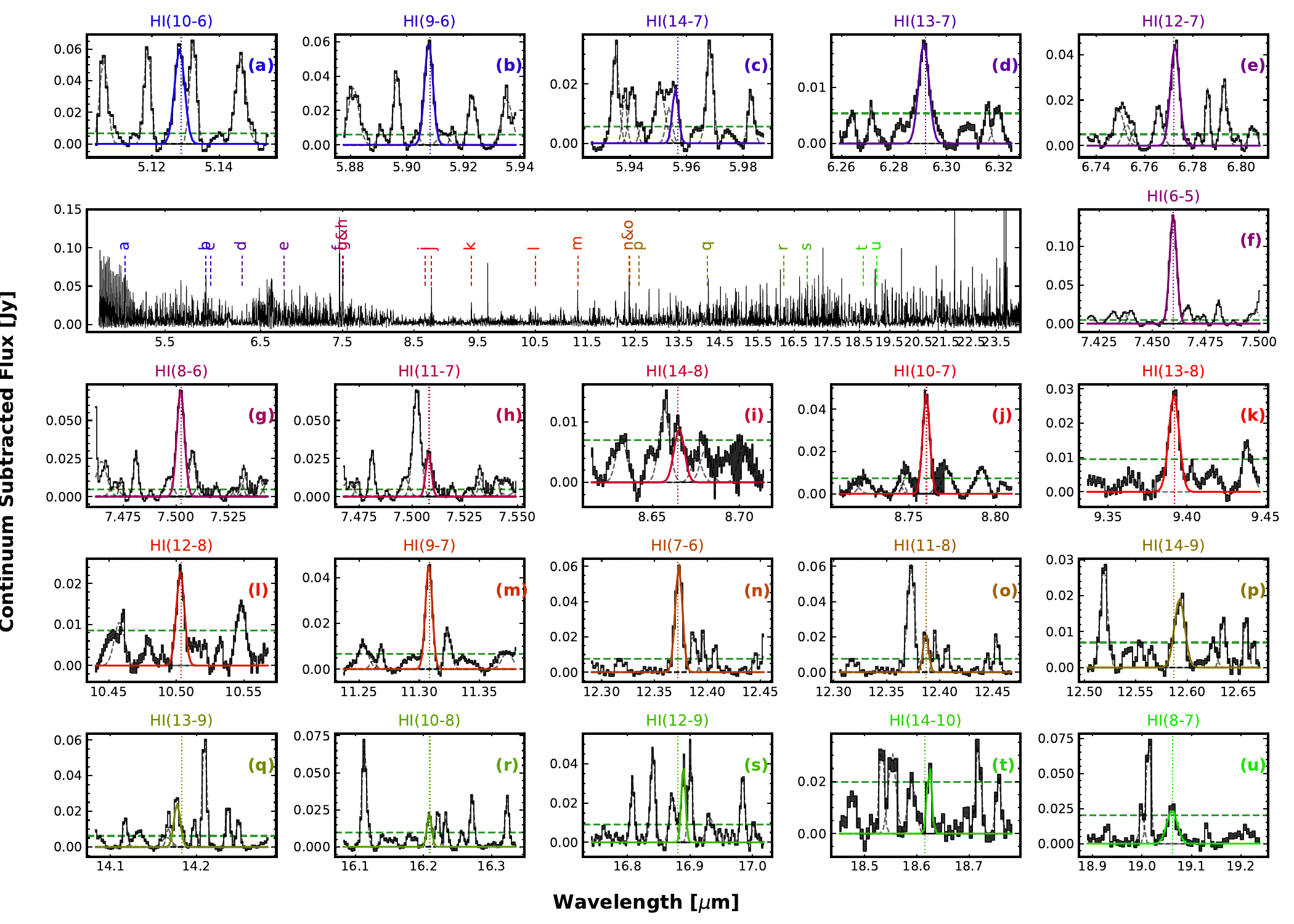}
    \caption{Continuum-subtracted \textit{JWST}/MIRI spectrum of FT~Tau. The main panel shows the full MIRI spectrum from 5 to 24~$\mu m$, with the positions of 22 H~\textsc{i} transitions marked. Sub-panels (a) through (u) show Gaussian fits (coloured curves) to individual H~\textsc{i} line profiles used in the analysis. The horizontal dashed green line indicates the $10 \times \mathrm{RMS}$ threshold used for line identification.}
    \label{fig1}
\end{figure*}

\section{Data inventory}
\label{sec:data}
\subsection{Sample and data reduction}

In this study, we utilized publicly available \textit{JWST}/MIRI data, specifically using the MRS mode \citep{Wells2015PASP..127..646W, Labiano2021AA...656A..57L,Wright2023PASP..135d8003W, Argyriou2023AA...675A.111A}. The sample consists of Class II and Class III sources spanning a range of spectral types from M-type to A-type stars. These data were obtained from the Mikulski Archive for Space Telescopes (MAST) and originate from several Cycle 1 general observer (GO) and guaranteed time observations (GTO) programs, including program IDs 1282 (PI: Thomas Henning, GTO; 45 sources), 1549 (PI: Klaus Pontoppidan, GTO; 3 sources), 1584 (PI: Colette Salyk, GO; 16 sources), 1640 (PI: Andrea Banzatti, GO; 8 sources), 1751 (PI: Melissa McClure, GO; 3 sources), and 2025 (PI: Karin Oberg, GO; 4 sources). Several studies based on these proposals have focused on the molecular inventory and outflows in these discs (\citealp{Grant2024AA...689A..85G, Temmink2024AA...689A.330T, Banzatti2025AJ....169..165B, Xie2023ApJ...959L..25X, Vlasblom2025AA...693A.278V, Salyk2025AJ....169..184S, Anderson2024ApJ...977..213A, Kanwar2024AA...689A.231K, Romero-Mirza2024ApJ...964...36R,Gasman2025A&A...694A.147G,Arabhavi2025ApJ...984L..62A,Schwarz2025ApJ...980..148S,Temmink2025arXiv250515237T,Perotti2025arXiv250411424P} and references therein). While these studies primarily targeted molecular inventories, here we focus on the H~\textsc{i} lines as tracers of accretion.

The sample comprises 79 PMS stars, predominantly spanning spectral types K7 to mid-M spectral type, known as T Tauri stars. The stellar parameters for the sample were retrieved from the literature, mainly from \citet{Manara2023ASPC..534..539M}. Stellar masses (\(M_*\)) range from 0.08 \(M_\odot\) to 2.18 \(M_\odot\). Notably, 68\% of the stars have \(M_* < 0.75 \, M_\odot\), highlighting the predominance of T Tauri stars in the sample. The few F- and A-type stars in the sample are primarily associated with debris discs or Class III objects, except for CD-25 11111B, an edge-on Herbig Ae star. The sample includes sources from nearby star-forming regions such as Taurus, Lupus, and Chamaeleon, with a median distance of $\sim$150 pc. Accretion luminosities (\(L_{\text{acc}}\)), from \citet{Manara2023ASPC..534..539M} and a few other references (see Table \ref{tab:big_table}), range from \(\sim10^{-4} \, L_\odot\) to \(\sim10 \, L_\odot\), while mass accretion rates (\(\dot{M}_{\text{acc}}\)) span from \(\sim10^{-11}\) to \(10^{-6} \, M_\odot \, \text{yr}^{-1}\). The sample spans five orders of magnitude in accretion rate at comparable distances, enabling a statistically robust comparison.

The raw \textit{JWST}/MIRI observations were uniformly processed with version 1.18.0 of the JWST calibration pipeline \citep{Bushouse2024zndo..14153298B}, using the Calibration Reference Data System (CRDS) context \texttt{jwst\_1364.pmap} and CRDS version 12.1.1. The reduction followed the standard three-stage workflow, similar to the reduction described in  \citet{federman2024investigating}, \citet{ narang2024discovery}, \citet{Neufeld2024ApJ...966L..22N}, and \citet{tyagi2025jwst}: detector-level calibration (Stage 1), spectral calibration (Stage 2), and cube building with spectral extraction (Stage 3), producing fully calibrated 3D spectral cubes and 1D extracted spectra for all 12 MIRI MRS sub-bands. To mitigate the effects of bad pixels, we implemented a pixel replacement algorithm using the minimum gradient (`\texttt{mingrad}') method in Stage 2. Additionally, a residual fringe correction was applied at the cube level. In Stage 3, spectral cubes were generated with the \texttt{output\_type=`band'} setting, automatically organizing data by spectral band. Outlier detection was applied (\texttt{outlier\_detection.skip = False}) with an 11$\times$11 pixel kernel (\texttt{kernel\_size = `11 11'}) and a 99.5\% detection threshold (\texttt{threshold\_percent = 99.5}), along with a secondary artifact check in the IFU data cubes (\texttt{ifu\_second\_check = True}). For the 1D spectral extraction, automatic source centring (\texttt{ifu\_autocen = True}) and residual fringe correction (\texttt{ifu\_rfcorr = True}) were enabled to improve the accuracy of the extracted spectra. The final data products consist of flux-calibrated 3D spectral cubes and corresponding 1D extracted spectra across all 12 MIRI sub-bands. While the 3D cubes were primarily used to investigate the spatial extent of the MIR H~\textsc{i} emission lines (Section \ref{spatial_sec2.2}), the subsequent analysis presented in this work was performed using the \texttt{x1d} 1D-spectroscopic data for each source. 

The MIRI MRS provides spectral resolving power ranging from approximately 1500 to 4000, depending on wavelength \citep{Jones2023MNRAS.523.2519J, pontoppidan2024ApJ...963..158P}. All sources were observed across the four MIRI MRS channels, covering wavelengths from 4.9~$\mu m$ to 27.9~$\mu m$. The spatial field of view varies across channels, ranging from approximately 3.2$^{\prime\prime}$~$\times$~3.7$^{\prime\prime}$ for Channel 1 Short to 6.6$^{\prime\prime}$~$\times$~7.7$^{\prime\prime}$ for Channel 4 Long. Extinction corrections were applied to the spectra using the extinction law from \citet{McClure2009ApJ...693L..81M}.\footnote{We also verified the results using the extinction curves from \citet{hensley2020detection} and \citet{kp5pontoppidan2024constrained}. Since the A$_V$ values are low for Class II sources, the choice of extinction curve does not significantly affect the line fluxes.} The $A_{\rm V}$ values were taken from \citet{Manara2023ASPC..534..539M} and other references. The stellar parameters for the sample are listed in Table \ref{tab:big_table}, along with the corresponding references.

\subsection{Spatial extent of H~\textsc{i} lines}
\label{spatial_sec2.2}

Several studies have reported extended emission in optical and NIR H~\textsc{i} lines originating from jets or shocked knots, which are often unrelated to accretion columns \citep{bajaj2025arXiv250323319B,Rayjets_2021, federman2024investigating,narang2024discovery,FloresRivera2023..670..A126, Beck2010ApJ...722.1360B, caratti016AA...589L...4C}. Thanks to its higher spatial resolution compared to \textit{Spitzer}, \textit{JWST} enables an assessment of the spatial extent of the H~\textsc{i} emission regions, particularly to evaluate whether large-scale jets contribute to the observed H~\textsc{i} emission.

Using the MIRI Integral Field Unit (IFU) 3D cubes, we generated continuum-subtracted line maps for the H~\textsc{i} (6--5) at 7.459 $\mu m$ and H~\textsc{i} (7--6) at 12.371 $\mu m$ transitions, using an extraction window of $\pm 2\,\Delta \lambda$ (where $\Delta \lambda = \lambda / R(\lambda)$), where \(R(\lambda) = 4603 - 128 \lambda + 10^{-7.4\lambda}\) \citep{Law2023AJ....166...45L}. Our analysis revealed that the spatial extent of both the H~\textsc{i} (6--5) and H~\textsc{i} (7--6) lines is confined within a radius of one PSF full width at half maximum (FWHM). This result remains valid even for sources where [Ne~\textsc{ii}] and [Fe~\textsc{ii}] lines exhibit highly collimated jets, suggesting that mid-infrared H~\textsc{i} emission is not spatially extended at the resolution of \textit{JWST} (see Appendix \ref{fig:contsub_linemap} for linemaps). We therefore conclude that mid-infrared H~\textsc{i} emission lines in our sample do not show contributions from spatially extended jets or outflows at the spatial resolution of \textit{JWST}. This supports the interpretation that, at least in Class II and III sources, the MIR H~\textsc{i} lines trace compact regions, likely associated with the inner accretion zone.

\section{Analysis and results}
\label{sec:analysis}
\subsection{Line detection}

\begin{figure}
    \centering
    \includegraphics[width=\linewidth]{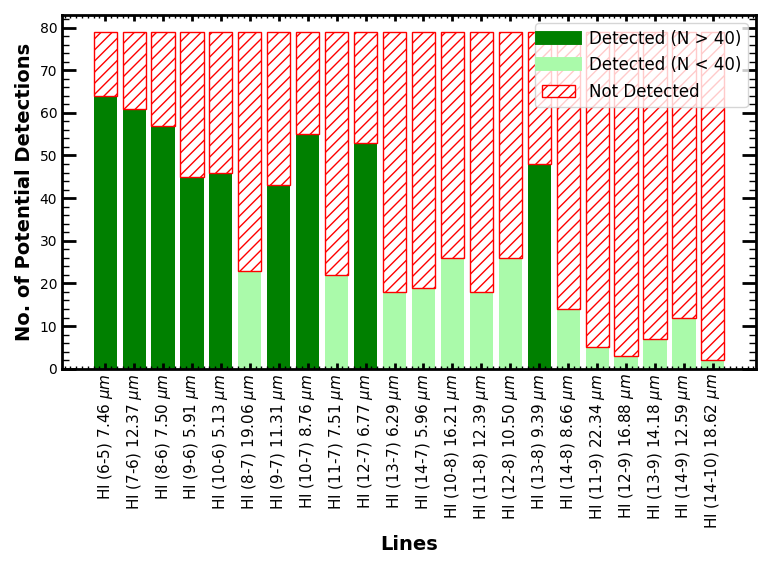}
    \caption{{H~\textsc{i} line detection statistics. Vertical stacked bar plot showing the detection statistics of H~\textsc{i} emission lines in our sample. Dark green bars denote lines detected in more than 40 sources, which are analysed in detail in this work. Light green bars correspond to lines detected in fewer than 40 sources, and red-hatched bars indicate the number of non-detections for each line.} 
}
    \label{fig:2stats}
\end{figure}

We analysed 22 H~\textsc{i} transitions with upper quantum numbers ($N_{\rm up}$) between 9 and 14, and lower quantum numbers ($N_{\rm low}$) between 5 and 11. The transitions span wavelengths from 5.0~$\mu m$ to 22.34~$\mu m$, covering the relatively unexplored H~\textsc{i} series, including Pfund ($N_{\rm low}=5$), Humphreys ($N_{\rm low}=6$), and higher-level transitions. Several additional H~\textsc{i} transitions with $N_{\rm up} > 14$ fall within the MIR range but are not included in this study, as they tend to be too faint for detection alongside the molecular features.

To identify and measure the fluxes of each H~\textsc{i} transition, we extracted spectral cut-outs spanning $30\times\, \Delta \lambda$ centred on the rest wavelength of each line. We then subtracted the local continuum by fitting the baseline using the \texttt{pybaselines} \citep{pybaselines} package, employing the \texttt{asymmetric least squares (ALS)} method, which effectively fits baselines containing both emission and absorption features. Peaks within this window were detected using the \texttt{scipy.find\_peaks} \citep{scipy2020SciPy-NMeth} function, applying a height threshold of $10\times\mathrm{RMS}$, where RMS is the root mean square of flux uncertainties derived from the errormaps within the 3D IFU data cubes. The \texttt{distance} parameter in \texttt{scipy.find\_peaks} was set to three spectral points to prevent closely spaced noise features from being misidentified as distinct peaks.

\begin{figure}
    \centering
    \includegraphics[width=\linewidth]{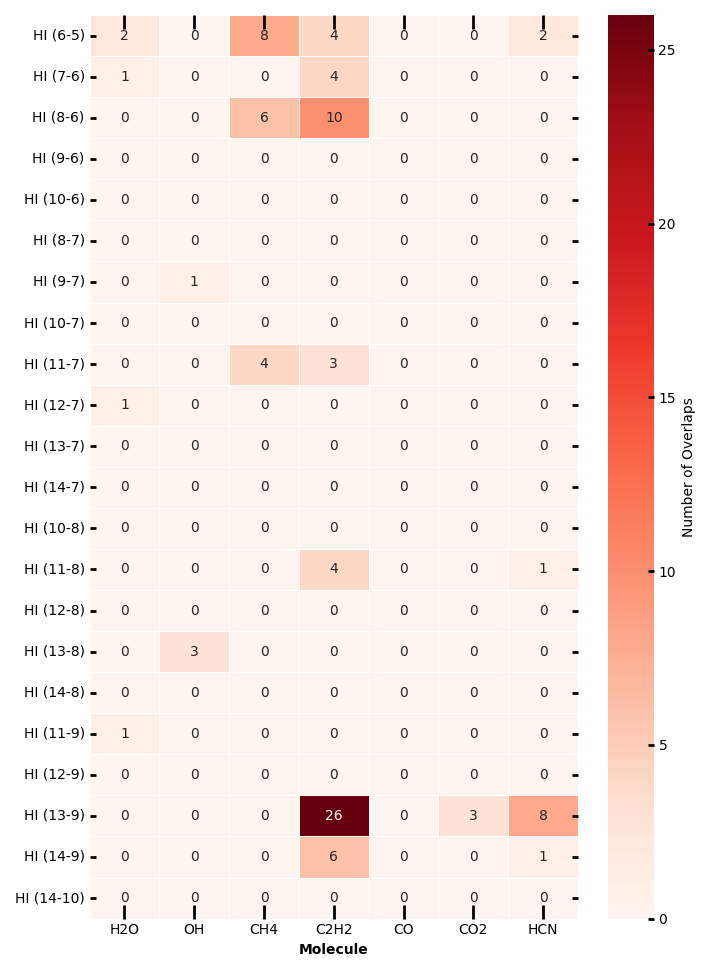}
    \caption{Heatmap showing the overlap of molecular transitions with the H~\textsc{i}. Each cell represents the number of molecular lines (E$_{\rm up} < 5000$~K) from species such as H$_2$O, HCN, CH$_4$, C$_2$H$_2$, CO, and CO$_2$ that fall within $\pm1\,\Delta\lambda$ of the rest wavelength of a given H~\textsc{i} transition. Darker colours indicate a higher number of overlapping molecular features, with H~\textsc{i} (13--9) exhibiting the strongest contamination, primarily from CH$_4$ and H$_2$O. Based on our analysis, lines such as H~\textsc{i} (9--6), H~\textsc{i} (10--6), and H~\textsc{i} (10--7) do not show significant molecular contamination.}
    \label{fig:2}
\end{figure}

All peaks above the $10\times\mathrm{RMS}$ threshold (dashed green line in sub-panels of Figure \ref{fig1}) were simultaneously fitted with Gaussian profiles using the \texttt{lmfit} \citep{lmfitnewville_2025_15014437} package. For each peak, a Gaussian model was initialized with amplitude, centre, and width (sigma) parameters, which were constrained to specific ranges to facilitate convergence given the large number of free parameters (three per Gaussian). The fitting window of $\pm15\,\Delta\lambda$ around the rest wavelength was adopted to balance the number of Gaussian components and the available spectral points. The most crucial constraint imposed on each Gaussian was that its sigma could vary between $0.8 \times \Delta \lambda$ and $2 \times \Delta \lambda$, while the centre was allowed to vary within $1 \times \Delta \lambda$ of the expected line centre. These constraints ensured that spurious features were not misidentified as target lines. Peaks were iteratively added to the model, with Gaussian parameters dynamically initialized to account for overlapping features or line broadening. The fitting procedure employed inverse variance weighting based on flux uncertainties to optimize the fits while minimizing the impact of noise.

The integrated area under each fitted Gaussian was adopted as the line flux for the corresponding H~\textsc{i} transition. Figure \ref{fig1} shows the Gaussian fits for each H~\textsc{i} transition in each sub-panel, along with the full continuum-subtracted spectrum of FT~Tau. An interactive step allowed user confirmation or rejection of each fit, providing an additional layer of quality control. This approach ensured a thorough and reliable methodology for line identification and flux measurement in complex spectra.

Line fluxes were converted to luminosities using distances from \textit{Gaia DR3}, or from \citet{Manara2023ASPC..534..539M} for sources lacking \textit{Gaia DR3} distances. It should be noted that the H~\textsc{i} line fluxes were not corrected for photospheric absorption. As reported by \citet{Salyk2025AJ....169..184S} and \citet{Tofflemire2025arXiv250408029T}, some higher-order H~\textsc{i} lines can appear in absorption in edge-on discs (e.g. MY Lup) or during low-accretion epochs (e.g. DQ Tau). Such corrections may be necessary for higher transitions ($N_{\rm up} > 12$) in the $N_{\rm low} = 6,7,8$ series for stars with $T_{\rm eff} > 7000$~K (primarily A- and B-type stars). Since our sample is dominated by low-mass stars, photospheric absorption corrections to H~\textsc{i} lines are beyond the scope of this work.

{Figure \ref{fig:2stats} shows the histogram of H~\textsc{i} line detections in our sample}. The H~\textsc{i} (6–5) line at 7.46~$\mu m$ is the most frequently detected, observed in 64 out of 79 sources (about 80\%). {The remaining 15 sources without H~\textsc{i} (6-5) detection are debris discs (i.e. low IR excess and very weak accretion) and edge-on discs (high-extinction) in the sample.} It is followed closely by the H~\textsc{i} (7–6) line at 12.37~$\mu m$ (61/79), H~\textsc{i} (8–6) at 7.50~$\mu m$ (57/79), and H~\textsc{i} (10–7) at 8.76~$\mu m$ (55/79). The H~\textsc{i} (8–7) line at 19.06~$\mu m$ is detected in only 23 out of 79 sources (23/79). As expected, detection rates decrease for transitions involving higher $N_{\rm up}$ {and $N_{\rm low}$ levels}. For example, the H~\textsc{i} (12–9) line at 16.88~$\mu m$ is detected in 3 out of 79 sources (3/79), and the H~\textsc{i} (14–10) line at 18.61~$\mu m$ is detected in only 2 out of 79 sources (2/79). 

However, this general trend is not always followed. Some lines, such as the H~\textsc{i} (12–7) line at 6.77~$\mu m$ (53/79) and the H~\textsc{i} (13–8) line at 9.39~$\mu m$ (48/79), show higher detection frequencies than those with lower $N_{\rm up}$ in the same series, such as the H~\textsc{i} (11–7) line at 7.508~$\mu m$ (22/79) and H~\textsc{i} (11–8) at 12.38~$\mu m$ (18/79). This anomaly may be due to unresolved molecular transitions, which cannot be fully separated at MIRI's spectral resolution. As a result, the apparent detection rates of these H~\textsc{i} lines may be artificially enhanced. Therefore, it is important to account for potential molecular contamination when reporting H~\textsc{i} detections.

\subsection{Contribution from molecular lines}

It is well known that the MIR spectra of Class~II discs are rich in emission features from various molecular species. Several studies have been conducted to characterize the molecular inventory of Class~II discs and to assess the material available for eventual planet formation (\citealp{Grant2024AA...689A..85G, Temmink2024AA...689A.330T, Banzatti2025AJ....169..165B} and others). Therefore, it is important to assess molecular transitions that may blend with H~\textsc{i} lines. Failure to account for molecular contributions can lead to overestimation of the H~\textsc{i} line fluxes. 

Using \textit{Spitzer} IRS short-high (SH) spectra, \citet{Rigliaco2015ApJ...801...31R} modelled the H$_2$O contribution to the H~\textsc{i} (7--6) and H~\textsc{i} (9--7) transitions following the method of \citet{Pontoppidan2010ApJ...720..887P}. They utilized the H$_2$O complexes around 15.17~$\mu m$ and 17.22~$\mu m$ to estimate and subtract the corresponding H$_2$O contribution from the H~\textsc{i} lines. Given the superior sensitivity and resolution of \textit{JWST}, detailed modelling of multiple molecular species is now feasible \citep[e.g.][]{arulanantham2025jdiscsurveylinkingphysics, Grant2024AA...689A..85G, Tabone2024AA...691A..11T, Temmink2024AA...689A.330T, Banzatti2025AJ....169..165B}. However, modelling all molecular species across the full sample of 79 discs is a time-intensive process and is beyond the scope of this work. Nevertheless, we provide an overview of the molecular lines that may blend with the H~\textsc{i} transitions analysed in this work.

We compiled the wavelengths of allowed molecular transitions (with upper state energy levels $E_{\rm up} < 5000$~K and $A_{ul} >$  1 $s^{-1}$) for abundant species including H$_2$O, CH$_4$, CO, CO$_2$, OH, C$_2$H$_2$, and HCN from the HITRAN database \citep{Gordon2022}. We cross-matched the line lists of these molecular species with the wavelengths of H~\textsc{i} transitions to identify potential overlaps. A molecular transition was considered overlapping if it fell within $1\Delta \lambda$ of any H~\textsc{i} transition. The potential overlaps are visualized as a heatmap in Figure \ref{fig:2}, with the number of overlaps annotated.

{Among these molecules, transitions of CH$_4$ and C$_2$H$_2$ show the most number of overlaps with H~\textsc{i} lines, followed by HCN and H$_2$O. However, CH$_4$ and C$_2$H$_2$ are not known to be prominent in Class II discs, particularly in objects earlier than M5 spectral type, at the wavelengths corresponding to the H~\textsc{i} lines. Therefore, they were not modelled in this work.} {For cooler, very low-mass stars (VLMS; later than M5), the contribution from hydrocarbon species such as CH$_4$ and C$_2$H$_2$ can be more significant. In such cases, their overlap with the H~\textsc{i} lines may need to be modelled carefully to avoid contamination. This has been highlighted in recent studies (e.g. \citealp{Arabhavi2025ApJ...984L..62A, Franceschi2024AA...687A..96F, Tabone2024AA...691A..11T}) that show how hydrocarbons can affect the mid-IR spectrum of cooler stars. }

{On the other hand, CO and CO$_2$ exhibit minimal overlap with the H~\textsc{i} lines and, as a result, were excluded from the modelling in this work.} The primary contaminant that requires careful modelling is H$_2$O. While the number of overlapping H$_2$O transitions is relatively small compared to some other molecules, H$_2$O emission lines are strong and present across the full MIRI wavelength range \citep{Banzatti2025AJ....169..165B}. Their broad presence and strength mean that even a relatively small number of overlapping lines can significantly affect the measured fluxes of the H~\textsc{i} lines. This is particularly true in the warmer regions of Class II discs, where H$_2$O is abundant and contributes prominently to the emission spectrum \citep[e.g.][and others]{Woitke2018A&A...618A..57W, Gasman2023A&A...679A.117G} Therefore, H$_2$O must be accurately accounted for, in the modelling of the H~\textsc{i} lines (see \citealp[e.g.][]{Banzatti2025AJ....169..165B, Rigliaco2015ApJ...801...31R, Tofflemire2025arXiv250408029T}).

{In the context of Class II discs, the H~\textsc{i} (6--5), H~\textsc{i} (7--6), and H~\textsc{i} (12--7) transitions are particularly affected by the presence of H$_2$O lines and therefore require explicit modelling to separate the contributions of H$_2$O from the intrinsic H~\textsc{i} emission. Additionally, the H~\textsc{i} (12--7) and H~\textsc{i} (13--8) transitions are affected by H$_2$O and OH, respectively. This explains their higher detection frequencies in comparison to other H~\textsc{i} transitions of the same H~\textsc{i} series, as these molecules could contribute to the observed emission, increasing the likelihood of their detection. Conversely, the H~\textsc{i} (14--9) and H~\textsc{i} (11--8) transitions were excluded from modelling due to their low detection frequencies.}

\begin{figure*}[h!]
    \centering
    \includegraphics[width=2.1\columnwidth]{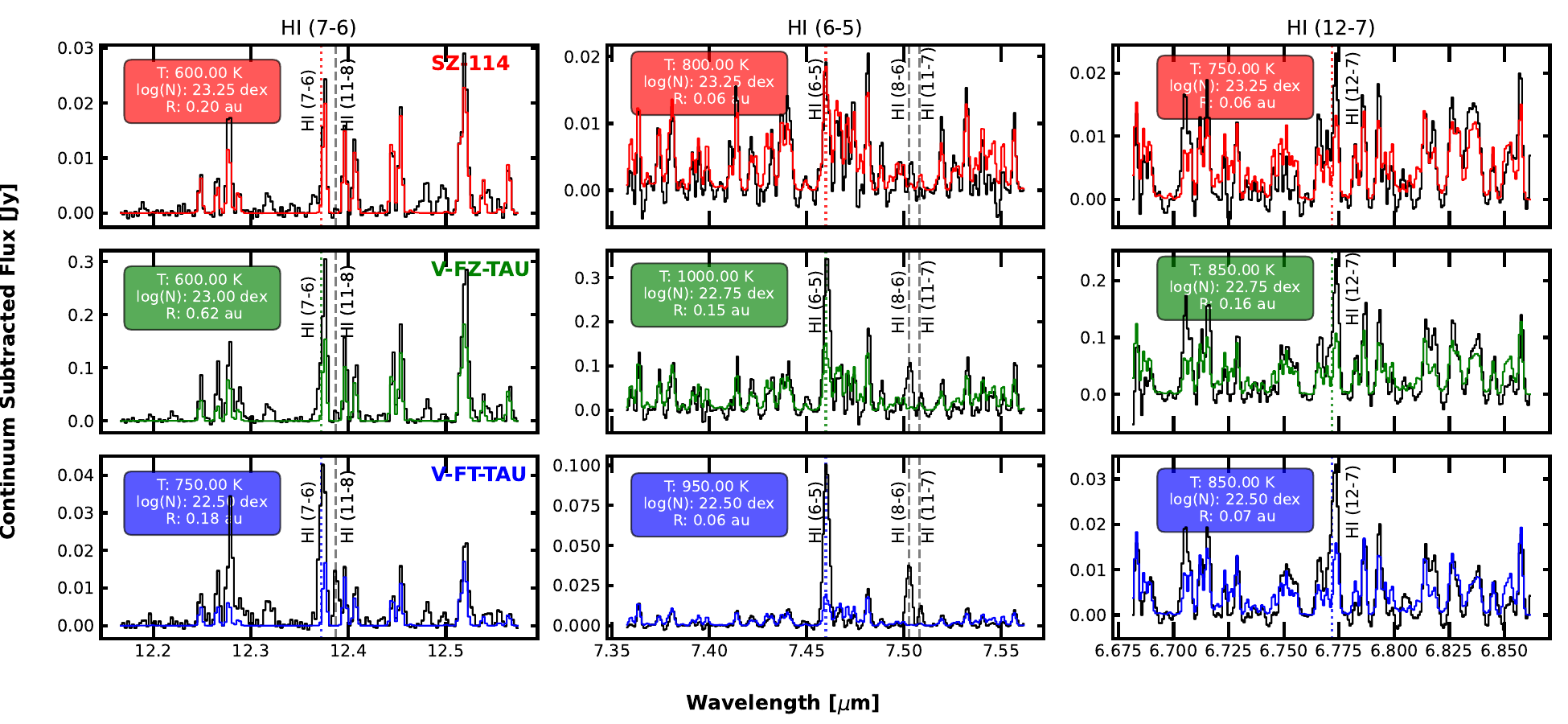}
    \caption{Representative best-fit LTE spectra for three Class~II discs. Sz~114 (top row), FZ~Tau (middle row), and FT~Tau (bottom row), showing varying levels of water contribution to H~\textsc{i} (7--6), H~\textsc{i} (6--5), and H~\textsc{i} (12--7) lines. In each panel, the black curve represents the continuum-subtracted \textit{JWST}/MIRI spectrum, while the coloured curve shows the best-fit LTE H$_2$O emission model. Inset boxes indicate the fitted physical parameters: temperature ($T$), column density ($\log(N)$), and emitting radius ($R$) for the water model. Other H~\textsc{i} lines are annotated in each of the sub-panels.}
    \label{fig:3}
\end{figure*}

As has been reported by \citet{Banzatti2025AJ....169..165B}, the H~\textsc{i} (10–7), (9–6), and (9–7) transitions do not exhibit significant molecular overlap that requires correction before flux measurements. We therefore restrict our correction to H$_2$O contamination for the H~\textsc{i} (12--7), (6--5), and (7--6) lines. For this work, modelling H$_2$O under the local thermodynamic equilibrium (LTE) assumption and subtracting its contribution is considered sufficient to ensure accurate H~\textsc{i} flux measurements.

As is illustrated in Figures 1--4 of \citet{Banzatti2025AJ....169..165B}, potential contributions from higher-$E_{\rm up}$ ($>$ 5000 K) transitions to the H~\textsc{i} lines are weak to negligible (as is also evident from Figure \ref{fig:3}, middle panels for the H~\textsc{i} (8--6) line) and are not considered as significant contaminants in our analysis. We further note that this selection criterion is used solely to identify which H~\textsc{i} lines require molecular contamination correction; the LTE modelling described in the following section applies no such $E_{\rm up}$ threshold and incorporates all molecular transitions available in the HITRAN \citep{Gordon2022} database.

\begin{figure*}
    \centering
    \includegraphics[width=2\columnwidth]{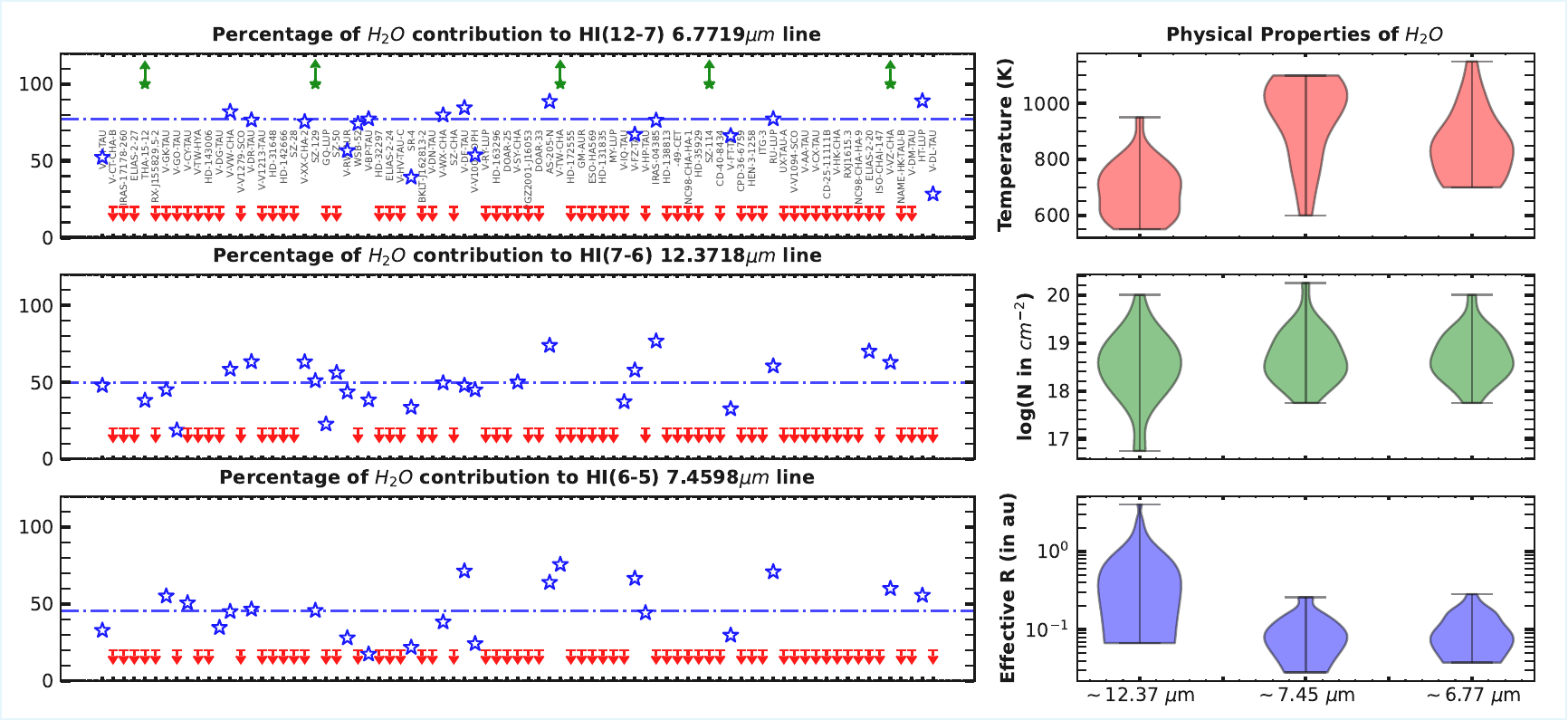}
    \caption{Contribution of H\textsubscript{2}O to 3 H~\textsc{i} lines. Left panels: Estimated percentage contribution of H$_2$O emission to the observed H~\textsc{i} transitions—H~\textsc{i} (12--7), H~\textsc{i} (7--6), and H~\textsc{i} (6--5)—across the Class~II disc sample. Blue stars indicate the fractional H$_2$O contribution for each source. Upward green arrows denote sources where the entire flux is attributable to H$_2$O, while downward red arrows mark sources for which H$_2$O modelling was not applied due to either dominant H~\textsc{i} flux or negligible H$_2$O emission. Right panels: Violin plots showing the distributions of best-fit physical parameters from LTE H$_2$O models for each wavelength region. The temperature and effective emitting radius exhibit an anti-correlation, consistent with compact hot inner disc emission dominating at shorter wavelengths. The H$_2$O column densities remain relatively constant across the different spectral regions.
}
    \label{fig:4}
\end{figure*}

\subsection{Removing H\textsubscript{2}O contamination from H~\textsc{i} lines}

To model the H\textsubscript{2}O lines around each H~\textsc{i} line, we performed an LTE 0D-slab modelling similar to \citet{Salyk2011ApJ...731..130S}. Assuming that the molecule is in LTE, each molecular line is treated as a Gaussian with thermal broadening corresponding to the {kinetic} temperature, \( T \). For a particular transition from upper level $u$ to lower level $l$, the line profile is given by

\begin{equation}
\phi_{ul}(v) = \frac{1}{\sigma_v \sqrt{2\pi}} e^{-\frac{v^2}{2\sigma_v^2}},
\end{equation}

where \(\sigma_v = \frac{\Delta v}{2 \sqrt{2 \ln{2}}}\), and \(\Delta v\) is the FWHM of the line. For a given column density, \(N\), and temperature, \(T\), the net optical depth, \(\tau_\nu\), arising from absorption and emission was calculated by summing up the contributions from all the overlapping transitions:

\begin{equation}
\tau_\nu = \sum_{u \rightarrow l} \frac{A_{ul} g_u \lambda_{ul}^3}{8\pi} \frac{N}{Z} \left( e^{-\frac{E_l}{k_B T}} - e^{-\frac{E_u}{k_B T}} \right) \phi_{ul}.
\end{equation}

Here:
\begin{itemize}
    \item \(A_{ul}\) is the Einstein A coefficient for the $u\rightarrow l$ transition,
    \item \(g_u\) is the upper state degeneracy,
    \item \(\lambda_{ul}\) is the transition wavelength,
    \item \(Z\) is the molecular partition function for the temperature T,
    \item \(E_u\) and \(E_l\) are the upper and lower energy levels of the transition, respectively,
    \item \(N\) is the column density of the molecule.
\end{itemize}

The flux, \(F_\nu\), is calculated as:
\begin{equation}
F_\nu = B_\nu(T) \left( 1 - e^{-\tau_\nu} \right) \Omega
\end{equation}

where \(B_\nu(T)\) is the Planck function at temperature \(T\), and \(\Omega = \frac{\pi R^2}{d^2}\) is the solid angle subtended by the emitting region, with \(R\) being the effective radius of the emitting area in AU, and \(d\) the distance in parsecs.

We utilized HITRAN spectroscopic database \citep{Gordon2022} and obtained the transition probabilities (\(A_{ul}\)), transition wavelengths (\(\lambda_{ul}\)), degeneracies ($g_u$), and the values of energy levels (\(E_u\), \(E_l\)) for all possible transitions of water in a given wavelength range. The inputs required to create synthetic LTE H\textsubscript{2}O spectra are temperature ($T$), column density ($N$), effective radius of the emitting area ($R$), and the distance to the source ($d$) to scale the spectrum.

We constructed a grid of LTE models over the parameter ranges:
\begin{itemize}
    \item Temperature: \(T \in [200, 2000] \, \text{K}\) (steps of \(50 \, \text{K}\)),
    \item Column density: \(\log_{10}(N(cm^{-2})) \in [12, 22]\) (steps of \(0.25 \, \text{dex}\)),
    \item Effective radius: \(\log_{10}(R) \in [-3, 1] \, \text{dex}  ~\text{(steps of } 0.125 \, \text{dex)}\).
\end{itemize}

For each parameter set, the partition function \(Z(T)\) was computed using total internal partition sums from supplemental data from the HITRAN database \citep{Gamache2021}. The Gaussian line profile \(\phi_{ul}(v)\) was constructed with a thermal line width of \(\Delta v = 4.7 \, \text{km/s}\), consistent with other works in the literature \citep[e.g.][]{Salyk2011ApJ...731..130S, Grant2024AA...689A..85G}. The optical depth \(\tau_\nu\) was then calculated over a velocity grid spanning \( \pm 5 \sigma_v\) around each transition. The resulting flux was evaluated over a fine grid of wavelength and subsequently convolved to match the JWST/MIRI spectral resolution, \(R(\lambda)\). Finally, the flux was resampled onto the observational wavelength grid to enable direct comparison with the observed spectra.

The best-fit model was taken as the one with the minimum \(\chi^2\) value, where \(\chi^2\) was calculated for each model using

\begin{equation}
\chi^2 = \frac{1}{\mathcal{N}} \sum_{i=1}^\mathcal{N} \left( \frac{F_{\text{data},i} - F_{\text{model},i}}{\sigma_i} \right)^2,
\end{equation}

where \(\mathcal{N}\) is the number of wavelength points, and \(\sigma_i\) represents the uncertainties in the observed flux. The best-fit H\textsubscript{2}O model was then visually verified. The fitting was not successful in some sources either because of strong H~\textsc{i} lines compared to molecular lines or low S/N of the spectra. The corresponding H\textsubscript{2}O line flux contribution from the best fit model was subtracted from the observed H~\textsc{i} line fluxes to obtain `true' H~\textsc{i} fluxes.

\begin{figure*}
    \centering
    \includegraphics[width=2\columnwidth]{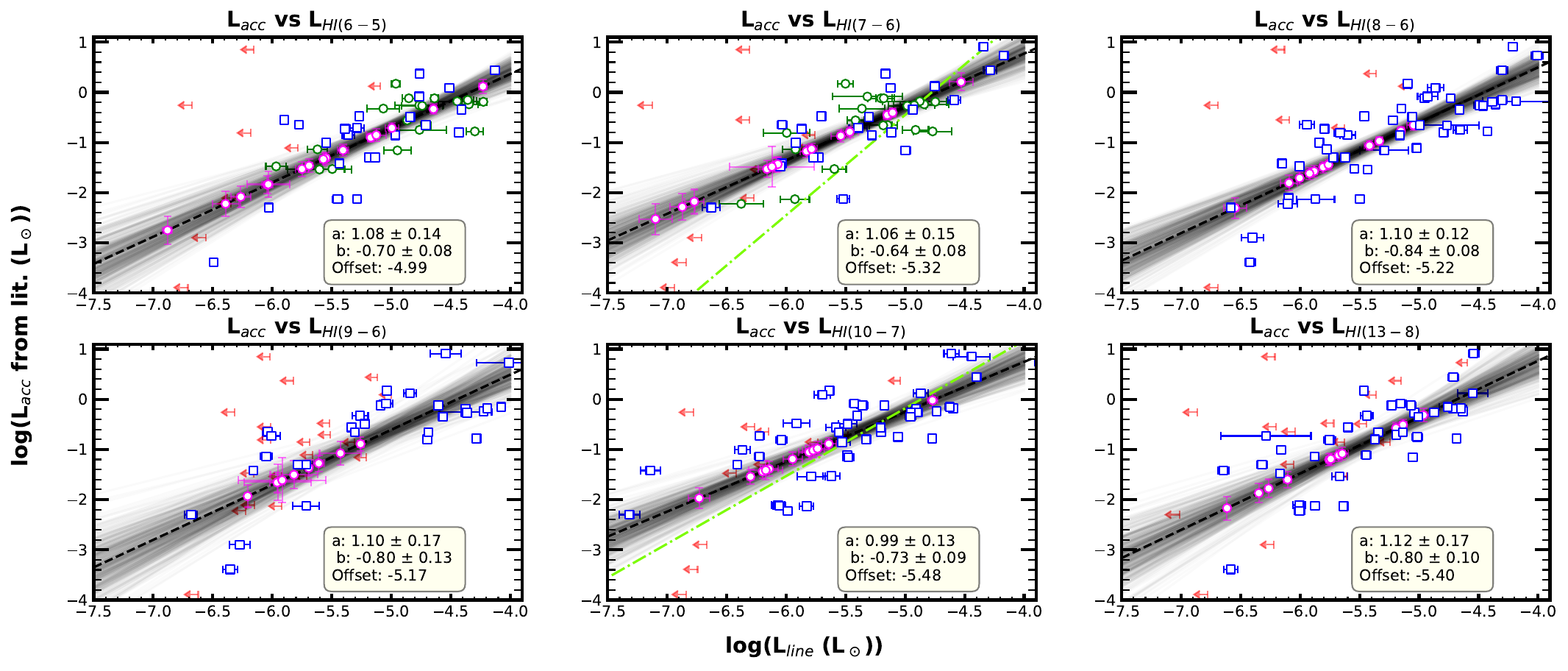}
    \caption{New and updated H~\textsc{i} empirical relations. Empirical correlations between accretion luminosity (log($L_{\rm acc}/L_{\odot}$)) and MIR H~\textsc{i} line luminosities (log($L_{\rm line}/L_{\odot}$)) for six transitions analysed in this work. Each panel displays the linear regression best-fit (dashed black line) with 1$\sigma$ and 3$\sigma$ confidence intervals (shaded regions). The best-fit slope and intercept are shown in the bottom-right corner of each panel and listed in Table~\ref{tab:2}. Blue open squares represent sources without correction for H$_2$O contamination, while green open circles correspond to sources where the H~\textsc{i} line fluxes have been corrected using LTE H$_2$O models. The pink circles represent the newly measured log(L$_{acc}$) values for sources that lack literature values, while the red arrows denote 3$\sigma$ upper limits for non-detections. The dash-dotted lines in the middle panels indicate the empirical relations reported by \citet{Tofflemire2025arXiv250408029T} for reference.}
    \label{fig:5}
\end{figure*}

Figure~\ref{fig:3} presents representative results for Sz~114, FZ~Tau, and FT~Tau, illustrating varying levels of H$_2$O contamination across different H~\textsc{i} transitions and a range of physical parameters resulting from the best-fit LTE models. The H$_2$O contribution is most pronounced for the H~\textsc{i} (12--7) transition, with a median contamination of approximately 75\% (Figure \ref{fig:4}, top left panel). This is expected, as H~\textsc{i} (12--7) is intrinsically weaker, yet its detection frequency is comparable to that of H~\textsc{i} (6--5), which is typically the strongest H~\textsc{i} line in the MIRI spectra. The median percentage of H$_2$O contribution is around 50\% for the H~\textsc{i} (7--6) and (6--5) transitions (see Figure \ref{fig:4}, left panels). This underscores the necessity of removing the water contribution before accurately measuring H~\textsc{i} fluxes for these lines.

Modelling of H$_2$O emission at three different wavelengths (12.37, 7.45, and 6.77~$\mu m$) also provides insights into the distribution of H$_2$O in T~Tauri discs. The right panels of Figure~\ref{fig:4} show the distributions of physical parameters from the best-fit H$_2$O LTE models. As previously noted by \citet{Banzatti2025AJ....169..165B} and others, the temperature of the H$_2$O emission increases with decreasing wavelength. This trend is mirrored in the effective radius, which decreases towards shorter wavelengths, consistent with hotter, more compact emission originating closer to the star. Specifically, H$_2$O emission at longer wavelengths ($\sim$12.37~$\mu m$) arises from cooler regions (\(\sim700\,\mathrm{K}\)) located around \(\sim1\,\mathrm{AU}\), while emission at shorter wavelengths ($\sim$6.77~$\mu m$) originates from hotter regions (\(\sim900\,\mathrm{K}\)) closer to the star (\(\sim0.10\,\mathrm{AU}\)). The column densities remain approximately constant across these regions, with most best-fit models yielding \(\log_{10}(N\,[\mathrm{cm}^{-2}]) = 18\)--19, in agreement with the results of \citet{Banzatti2025AJ....169..165B}.

Without these corrections, MIR-derived accretion rates based on H~\textsc{i} lines would be significantly affected by H$_2$O emission. This further highlights the utility of uncontaminated transitions such as H~\textsc{i} (8--6) and (10--7) as accretion indicators, even though they may be weaker than H~\textsc{i} (6--5) and (7--6).

\subsection{New and updated empirical relations to estimate L$_{acc}$}
\label{sec:empirical}

Most H~\textsc{i} lines have been widely used as proxies for estimating accretion rates in large samples of PMS stars. However, low $N_{up}$ H~\textsc{i} transitions such as H$\alpha$, Pa$\alpha$, and even Br$\alpha$ can be affected by additional contributions from jets or shocked knots \citep{Eisner2015MNRAS.447..202E, Muzerolle2001ApJ...550..944M, bajaj2025arXiv250323319B}. Moreover, high line-of-sight extinction at optical and NIR wavelengths limits our ability to probe more embedded sources. Higher-order H~\textsc{i} transitions in the MIR offer a way to overcome this limitation, potentially making them more reliable accretion indicators in the \textit{JWST} era. As discussed in Section~\ref{spatial_sec2.2}, MIR H~\textsc{i} lines are not spatially extended, in our sample of Class II sources, exhibiting collimated [Ne~\textsc{ii}] and [Fe~\textsc{ii}] jets. Therefore, MIR H~\textsc{i} lines offer a means to probe the innermost regions of circumstellar discs, while minimizing contributions from jets and outflows.

The sub-panels of Figure~\ref{fig:5} shows the best-fit empirical relations for converting H~\textsc{i} line fluxes for six different transitions (selected based on higher detection rate and low molecular contamination) into accretion luminosities, compiled from literature values (primarily from \citealp{Manara2023ASPC..534..539M}, refer to Table \ref{tab:big_table} for references). We exclude the H~\textsc{i} (12--7) transition despite its high detection frequency due to significant H$_2$O contamination. {H~\textsc{i} (10--6) is avoided as it is present among the CO fundamental line forest, which could have complex emission/absorption morphologies based on the nature and inclination of the disc \citep{Banzatti2022AJ....163..174B}. Between H~\textsc{i} (9--7) and (13--8) transitions, we chose to use the (13--8) transition as shorter wavelength ($\sim$ 9 $\mu m$) OH lines are much weaker compared to longer-wavelength OH lines. We include this line in the analysis as the OH emission could originate from UV photo-dissociation of H$_2$O and also can serve as a proxy for UV excess from accretion \citep{Tabone2024AA...691A..11T, Zannese2023AA...671A..41Z, Neufeld2024ApJ...966L..22N}}.  Table~\ref{tab:2} reports the best-fit parameters for these six transitions, along with their uncertainties.  In each panel of Figure \ref{fig:5}, green circles represent the corrected H~\textsc{i} line fluxes for sources where H$_2$O contamination has been removed, while blue squares represent measurements where H$_2$O correction could not be applied, either due to strong H~\textsc{i} emission or low S/N of the spectra.

We adopt the following linear relation, 

\[
\log\left(\frac{L_{\rm acc}}{L_{\odot}}\right) = a \times \left(\log\left(\frac{L_{\rm line}}{L_{\odot}}\right) - \text{offset}\right) + b, 
\]

to create an empirical relation for MIR H~\textsc{i} lines. We introduced an offset to reduce the covariance between the slope ($a$) and intercept ($b$), following the approach of \citet{Tofflemire2025arXiv250408029T}. The offset for each line was set to the mean of the respective $\log\left(L_{\rm line}/L_{\odot}\right)$ values for our sample \footnote{See \citet{draper1998applied} and other such references.}. The parameters $a$, $b$, and the offset were estimated using the \texttt{linmix} package, which applies a Bayesian framework to fit the relation while properly propagating uncertainties in both $x$ and $y$. Magenta markers represent sources for which accretion luminosities were not available in the literature; for these, $L_{\rm acc}$ was estimated from the observed MIR H~\textsc{i} line luminosities using the fitted relation (see Tab. \ref{tab:acc_estimates}). 

\begin{figure*}
    \centering
    \includegraphics[width=\linewidth]{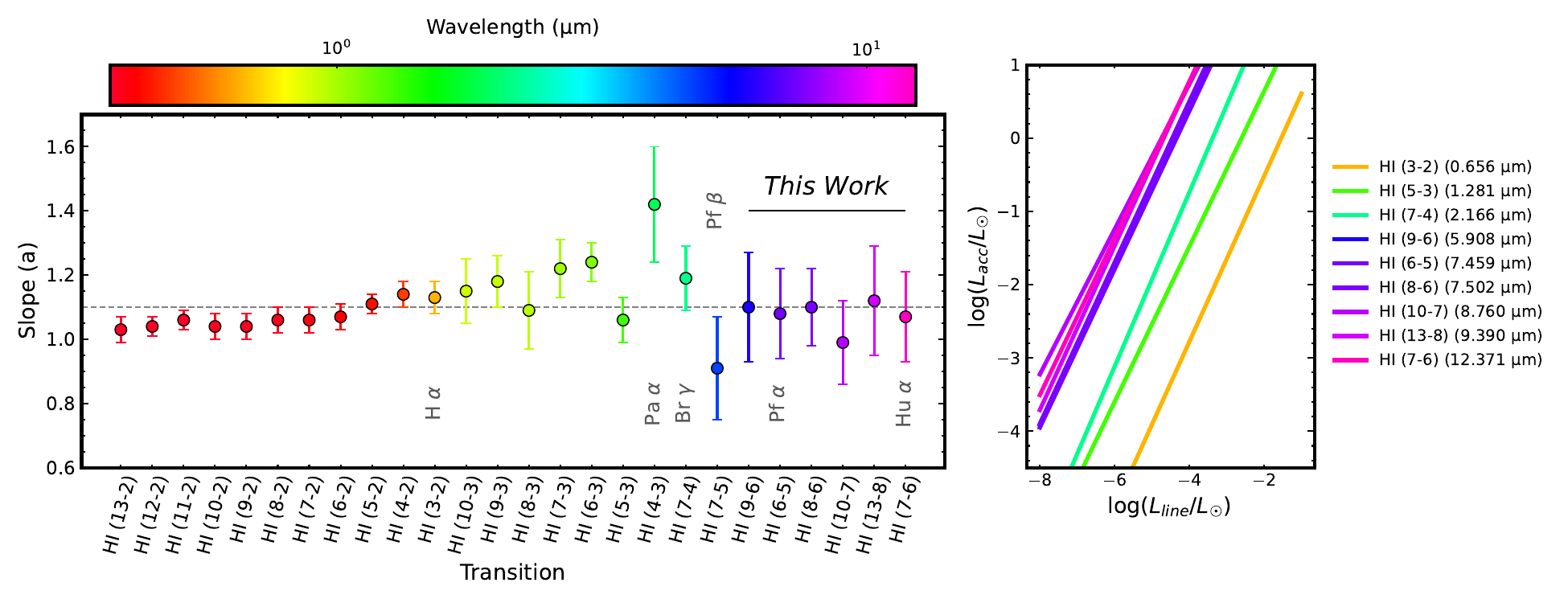}
    \caption{{Comparison of different H~\textsc{i} empirical relations. (Left) Estimated empirical slopes of optical/NIR H~\textsc{i} lines from the literature \citep{Alcala2017AA...600A..20A, Rogers2024AA...684L...8R, Salyk2013ApJ...769...21S} and mid-infrared H~\textsc{i} lines from this work. The colour bar and marker colours indicate the wavelength of each H~\textsc{i} transition. The median slope of the distribution is $\sim$1.1, shown with a dashed line.} The slope of the Pa$\alpha$ empirical relation \citep{Rogers2024AA...684L...8R} is slightly offset from other H~\textsc{i} lines, as it was calibrated using \textit{JWST} spectra of young stars in the Large Magellanic Cloud (LMC), which has sub-solar metallicity. This lower metallicity likely affects the slope. {(Right) Comparison of empirical relations between accretion luminosity ($\log L_{\mathrm{acc}}$) and line luminosity ($\log L_{\mathrm{line}}$) for representative H~\textsc{i} transitions from optical to mid-infrared wavelengths. The slopes are consistent across this wide wavelength range.}}
    \label{fig:6}
\end{figure*}

\begin{table}[h!]
\caption{New and updated H~\textsc{i} empirical relations}
\label{tab:2}
\centering
\begin{tabular}{lcccccc}
\hline\hline
Line & $\lambda$ ($\mu$m) & $a$ & $a_{\text{err}}$ & $b$ & $b_{\text{err}}$ & Offset \\
\hline
H~\textsc{i}(9-6)  & 5.908  & 1.10  & 0.17  & -0.80 & 0.13 & -5.17 \\
H~\textsc{i}(6-5)  & 7.459  & 1.08  & 0.14  & -0.70 & 0.08 & -4.99 \\
H~\textsc{i}(8-6)  & 7.502  & 1.10  & 0.12  & -0.84 & 0.08 & -5.22 \\
H~\textsc{i}(10-7) & 8.760  & 1.00  & 0.13  & -0.73 & 0.09 & -5.48 \\
H~\textsc{i}(13-8) & 9.390  & 1.12  & 0.17  & -0.80 & 0.10 & -5.40 \\
H~\textsc{i}(7-6)  & 12.371 & 1.07  & 0.14  & -0.64 & 0.08 & -5.32 \\
\hline
\end{tabular}
\tablefoot{Fit parameters for H~\textsc{i} lines of the form $\log(L_{\rm acc}/L_{\odot}) = a \times [\log(L_{\rm line}/L_{\odot}) - \text{offset}] + b$. An offset is introduced to reduce the covariance between the slope ($a$) and the y-intercept ($b$), which can affect the precision of their estimates.}
\end{table}

\subsubsection{Effect of accretion variability}

An important caveat to note is that the literature measurements of log($L_{\rm acc}$) are not contemporaneous with the \textit{JWST} observations. The scatter observed in the correlation between literature log($L_{\rm acc}$) values and the MIR H~\textsc{i} line luminosities (log $L_{\rm line}$) can be explained by the intrinsic accretion variability of YSOs. The median accretion variability in T Tauri stars is reported to be a factor of 2-3 over timescales of days to weeks \citep[and others]{Fischer2023ASPC..534..355F, Flaischlen2022AA...666A..55F, Costigan2012MNRAS.427.1344C, Mendigutia2011AA...535A..99M, Venuti2014AA570A82}. It is therefore important to assess how this variability impacts the empirical relations derived in this work.

The most effective way to mitigate this issue is to obtain (near-)simultaneous measurements of UV excess or other accretion tracers alongside the MIR spectra. This approach has been adopted by \citet{Tofflemire2025arXiv250408029T}, who performed coordinated \textit{JWST}, ground-based VLT/X-Shooter, and Las Cumbres Observatory photometric observations for the known binary DQ~Tau. However, extending such coordinated campaigns to statistically large samples would require substantial observational effort.

We assessed the impact of this variability by performing a Monte Carlo simulation test. For the observations shown in Figure~\ref{fig:5}, we introduced a random $\pm$50\% scatter to the observed MIR H~\textsc{i} line fluxes and rerun the linear regression fits for 1000 iterations. Figure~\ref{fig:monte_carlo_dist} (in appendix) shows the distributions of the resulting slope, intercept, and offset values as violin plots for each line. Each sub-panel also shows the reported slope, intercept, and offset values from Table~\ref{tab:2}, overplotted on the violin distributions. The results indicate that variability of $\pm$50\% has a minimal effect on the reported correlation slopes, which remain within the estimated uncertainties. These results provide confidence that the effect of variability on the empirical relations is minor and well within the reported uncertainties.

An alternative way to reduce the observed scatter due to variability is to adopt H~\textsc{i} (8--6), which is free of molecular contamination (Section~3.2), as the reference accretion tracer. Using log($L_{\rm acc\ (8-6)}$), we calibrated the remaining H~\textsc{i} lines to reduce the uncertainties in the derived empirical relations. The results are presented in Figure~\ref{fig:lacc_8-6} (in appendix), where it is evident that the scatter in the log($L_{\rm acc}$) vs log($L_{\rm line}$) relations is significantly reduced compared to Figure~\ref{fig:5}, as the influence of non-simultaneous measurements has been mitigated. The best-fit empirical relations from this analysis have much lower uncertainties compared to those derived using literature values. However, this approach should be interpreted with caution, as log($L_{\rm acc\ (8-6)}$) itself was originally estimated using literature values of log($L_{\rm acc}$), and thus introduces a `feedback loop' into the uncertainty estimates.

The newly derived accretion luminosities and mass accretion rates (including upper limits) obtained from H~\textsc{i} (6--5), H~\textsc{i} (7--6), H~\textsc{i} (8--6), and H~\textsc{i} (10--7) for our sample are given in Table~\ref{tab:acc_estimates}. The accretion luminosities measured from different H~\textsc{i} lines are in good agreement within the quoted uncertainties (Fig. \ref{fig:lacc_cons}). Under the assumption that H~\textsc{i} (8--6) provides the most reliable accretion estimate in MIRI, we find that log($L_{\rm acc}$) values derived from other lines are consistent with those from H~\textsc{i} (8--6), with a mean scatter of 0.2 dex. This further supports the robustness of the empirical relations and the newly derived accretion luminosities presented in this work.

\begin{figure*}
    \centering
    \includegraphics[width=2\columnwidth]{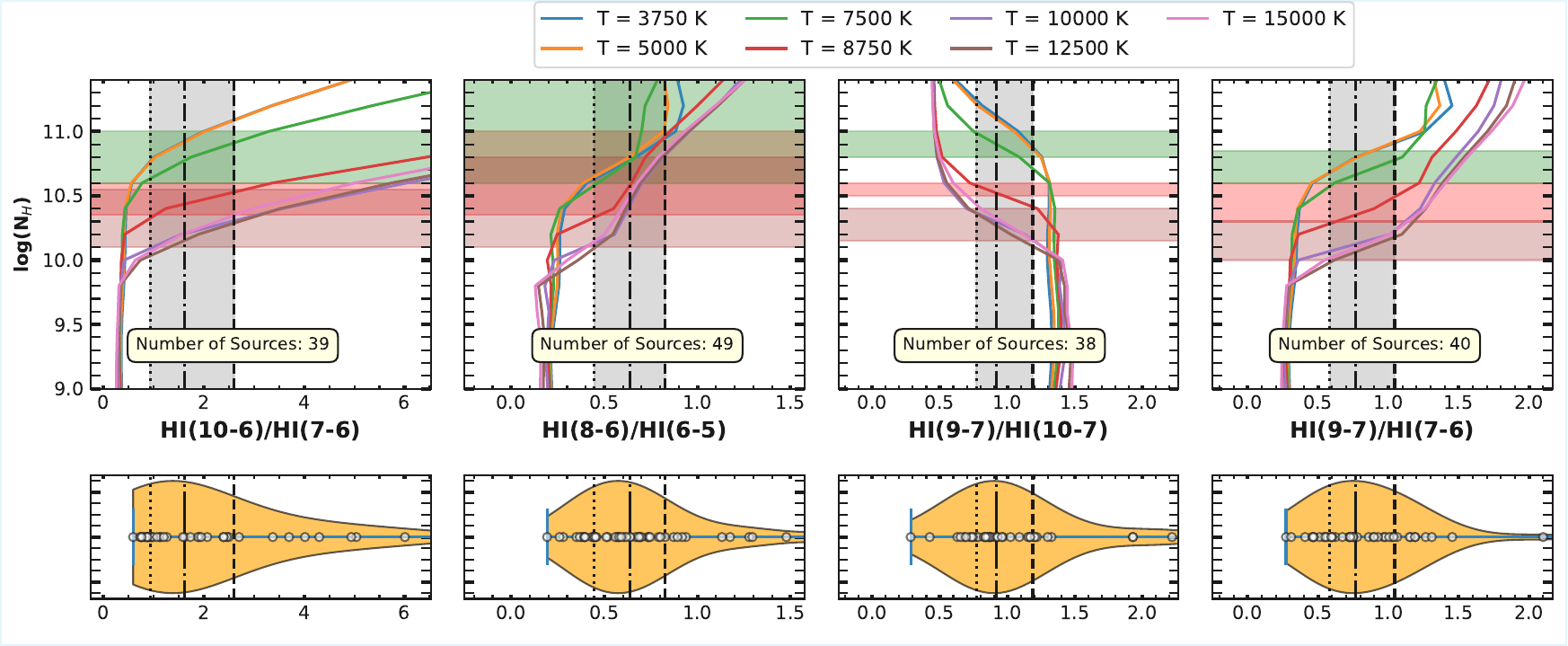}
    \caption{Comparison of observed mid-infrared (MIR) H~\textsc{i} line ratios with model predictions from \citetalias{Kwan2011MNRAS.411.2383K}. Top panels show theoretical line ratio curves as a function of hydrogen density ($n_{\mathrm{H}}$) for different gas temperatures, with coloured curves corresponding to temperatures ranging from 3750\,K to 15,000\,K. Vertical lines represent the observed median (dash-dotted), 25th percentile (dotted), and 75th percentile (dashed) values for each ratio. The bottom panels display horizontal violin plots of the observed line ratios from our sample, with circles indicating individual ratio values. {In each of the top panels, the shaded regions show the inferred range of $n_H$ values for three different temperature grids, with the colours corresponding to the respective temperatures (green for T=7500K, red for T=8750, and brown for T=12500K).} These comparisons demonstrate the sensitivity of certain MIR H~\textsc{i} line ratios (e.g. H~\textsc{i} (10--6)/H~\textsc{i} (7--6), H~\textsc{i} (9--7)/H~\textsc{i} (10--7)) to hydrogen densities $\gtrsim 10^{10.2}$~cm$^{-3}$.}
    \label{fig:7}
\end{figure*}

\begin{figure*}[h!]
    \centering
    \includegraphics[width=2.1\columnwidth]{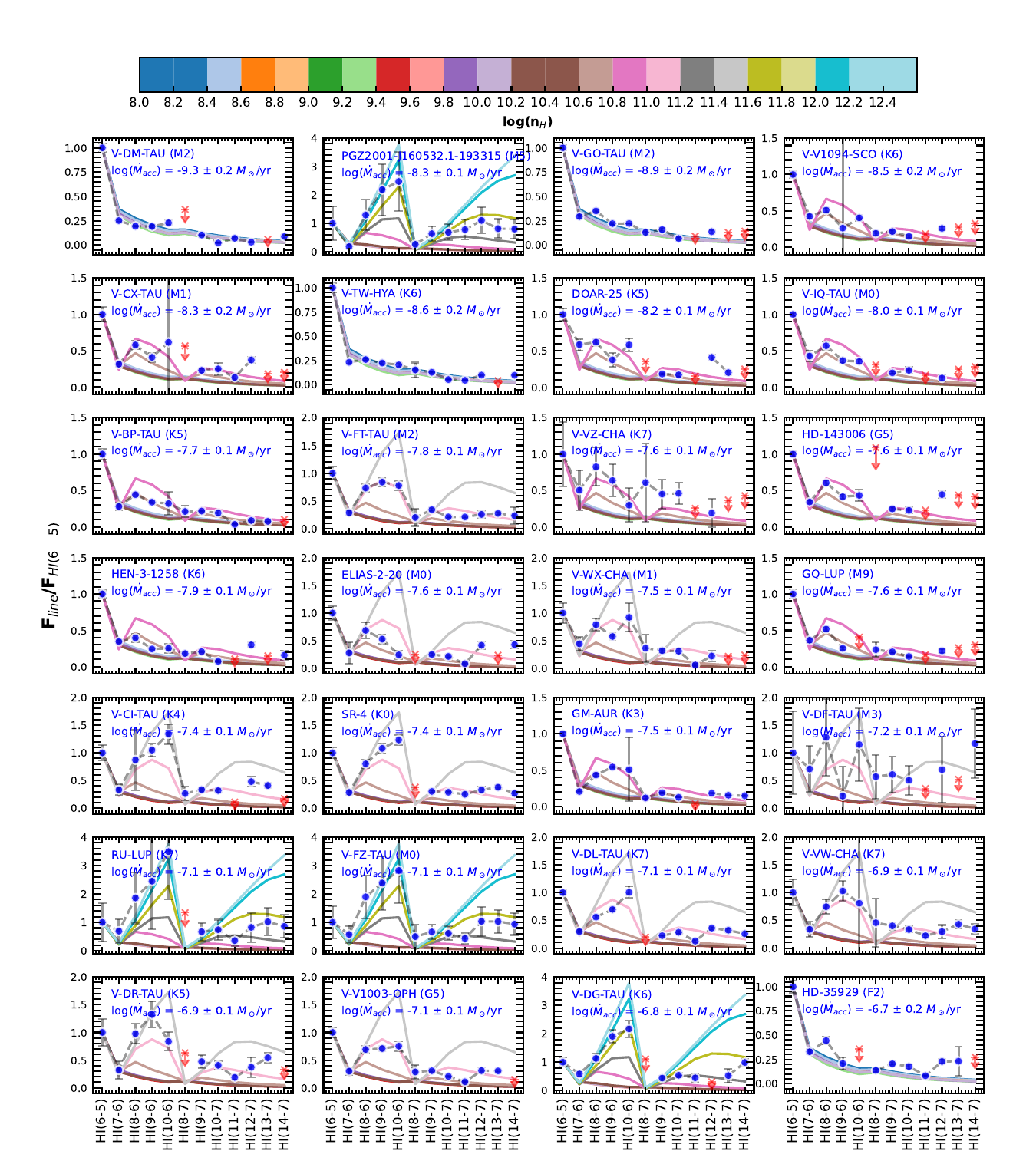}
    \caption{Comparison of observed and modelled H~\textsc{i} line decrements. The ratios are normalized to H~\textsc{i} (6--5) for a sample of Class~II sources and overlaid with model predictions from \citetalias{Kwan2011MNRAS.411.2383K} at a fixed temperature of 8750\,K. Each panel corresponds to an individual source, with blue-filled circles representing the observed line ratios and downward red arrows denoting upper limits for non-detections. Coloured model curves indicate the predicted line ratios for varying hydrogen densities shown in the colour bar above.}
    \label{fig:8}
\end{figure*}

\subsubsection{Comparison to other H~\textsc{i} relations}

{
By using \textit{Spitzer} IRS spectra, \citet{Rigliaco2015ApJ...801...31R} reported a strong positive correlation between H~\textsc{i} (7--6) line luminosity and accretion luminosity for a sample of 114 protoplanetary discs spanning various evolutionary stages. {As was previously reported by \citet{Rigliaco2015ApJ...801...31R},} owing to the low spectral resolution of \textit{Spitzer}, blending of H~\textsc{i} (7--6) with nearby contaminants, particularly with H~\textsc{i} (11--8) at 12.38~$\mu m$, is significant and cannot be resolved. They reported an empirical relation with a correlation slope of $\sim$2 for H~\textsc{i} (7--6) (blended with  H~\textsc{i} (11-8) line), which was detected in 46 out of 114 sources. The sensitivity and wavelength coverage of \textit{Spitzer} did not permit investigation of additional H~\textsc{i} transitions in the MIR. }

Through a highly coordinated, multi-telescope effort, \citet{Tofflemire2025arXiv250408029T} derived empirical relations for three MIR H~\textsc{i} lines: 7--6, 10--7, and 8--7 (shown in Figure~\ref{fig:5}, middle panels, as dashed bright green lines) using DQ Tau. Among these, they concluded that H~\textsc{i} (7--6) is affected by rotational H$_2$O emission, while H~\textsc{i} (10--7) and H~\textsc{i} (8--7) appear uncontaminated. Consistent with their findings, we also report that H~\textsc{i} (7--6) is not a reliable accretion indicator unless H$_2$O contamination is carefully modelled and removed, introducing additional complexity to its use as an accretion diagnostic.
The middle panels of Figure~\ref{fig:5} (dashed bright green lines) show the empirical relations for H~\textsc{i} (10--7) and (7--6) as reported by \citet{Tofflemire2025arXiv250408029T}, the latter being broadly consistent with the earlier relation reported by \citet{Rigliaco2015ApJ...801...31R}. The H~\textsc{i} (10--7) relation shows good agreement with our empirical calibration, within the estimated uncertainties. In contrast, the H~\textsc{i} (7--6) relation deviates significantly from our results. 

{The difference between H~\textsc{i} (7--6) relation of ours and \citet{Tofflemire2025arXiv250408029T} may largely stem from the choice of the sources used: their relations are based primarily on DQ Tau, whereas our analysis provides statistically robust correlations derived from a larger sample. Furthermore, our L$_{acc}$ measurements derived from different HI lines are mutually consistent (as is also shown in Fig. \ref{fig:lacc_cons}), in contrast to the results of \citet{Tofflemire2025arXiv250408029T}, where the L$_{acc}$ values inferred from HI (7–6) and (10–7) are not consistent with each other (see Table 4 of \citealp{Tofflemire2025arXiv250408029T}). Also, as noted by \citet{nayak2024ApJ...963...94N}, applying the H~\textsc{i} (7--6) relation from \citet{Rigliaco2015ApJ...801...31R} (and hence \citealp{Tofflemire2025arXiv250408029T}) results in unphysical accretion rates for their sources. This suggests that our empirical relation, with a slope close to unity, may better represent the true scaling of MIR H~\textsc{i} lines with accretion luminosity.} Given these discrepancies, we caution against the use of H~\textsc{i} (7--6) for accretion rate determinations. Instead, we recommend the use of uncontaminated lines such as H~\textsc{i} (8--6) at 7.502 $\mu m$, (10--7) at 8.760 $\mu m$, and (9--6) at 5.908 $\mu m$, which avoid complications introduced by H$_2$O contamination.

{In addition, we also compare the MIR relations to optical and NIR relations from the literature.} 
Figure~\ref{fig:6} compares the empirical slope for all known H~\textsc{i} lines reported in the literature. The empirical relations derived for the new MIR H~\textsc{i} lines in this study are in good agreement with the established optical and NIR relations, confirming their reliability. We do not observe the change in slope with increasing wavelength as reported by \citet{Tofflemire2025arXiv250408029T} (Fig. \ref{fig:6}, left). Notably, all slopes listed in Table \ref{tab:2} are consistent within the error bars, with a value close to unity, which aligns with well-established optical/NIR H~\textsc{i} relations (Fig. \ref{fig:6}, right).

\subsection{Estimating the physical conditions of the H~\textsc{i} emitting gas}
\label{sec:physcond}

Hydrogen decrement analysis has been widely employed to estimate the physical conditions of the emission regions in YSOs. Initially, the Case~B recombination model was used as an approximation to constrain parameters such as density and temperature \citep{Nisini2004AA...421..187N, Bary2008ApJ...687..376B, Kraus2012ApJ...745...19K, Whelan2014AA...570A..59W}. However, this model, which assumes optically thin gas, radiative ionization, and recombination, fails to capture the complexities of the inner disc environments surrounding YSOs. Observations of H~\textsc{i} line ratios have consistently shown discrepancies with Case~B predictions, necessitating the adoption of local line excitation models, such as those proposed by \citet[hereafter KF11]{Kwan2011MNRAS.411.2383K}, which incorporate collisional effects, optical depth variations, and local physical conditions.

Using X-Shooter spectroscopy of {35 low-mass YSOs in Lupus star-forming region}, \citet{Antoniucci2017AA...599A.105A} conducted a detailed study of Balmer and Paschen decrements in T~Tauri stars, identifying two distinct populations based on decrement shapes and line profiles. Stars with narrow, symmetric Balmer profiles exhibited decrements consistent with $n_{\rm H} \sim 10^9$~cm$^{-3}$ at $T = 5000$--15,000~K, indicative of optically thin emission. In contrast, stars with broader, multi-peaked profiles displayed L-shaped Balmer decrements corresponding to $n_{\rm H} > 10^{11}$~cm$^{-3}$, characteristic of optically thick emission from denser regions. These trends showed a strong correlation with accretion rates, suggesting that higher accretion activity enhances the density of gas in magnetospheric accretion flows.

The Paschen series, extensively examined by \citet{Edwards2013ApJ...778..148E} {IRTF/SpeX spectra of 16 YSOs in the Tau-Aur star-forming region}, provided further insights into these environments. By analysing Paschen decrements in T Tauri stars, they applied the \citetalias{Kwan2011MNRAS.411.2383K} models to constrain hydrogen densities to $n_H = 2 \times 10^{10}$ cm$^{-3}$ to $2 \times 10^{11}$ cm$^{-3}$. This range represents the transition from optically thin to optically thick regimes, where collisional excitation and level population build-up significantly affect emission properties. Furthermore, \citet{Edwards2013ApJ...778..148E} reported that higher accretion rates corresponded to increased densities. However, temperature estimates remained degenerate in these models, as the observed line ratios were sensitive to both density and temperature. The shift in Paschen decrement behaviour at these densities underscored the limitations of Case B models, which yielded inconsistent parameter estimates.

The Brackett series further reinforced these findings. Studies of Br$\gamma$ and higher-order transitions have indicated similar density estimates, confirming that these lines primarily trace accretion-dominated gas. \citet{Antoniucci2017AA...599A.105A} found that Brackett decrements, like Balmer and Paschen decrements, closely followed \citetalias{Kwan2011MNRAS.411.2383K} model predictions for dense, optically thick accretion flows. These results emphasized the necessity of incorporating local excitation effects, where collisional processes and optical depth variations significantly impact the observed line ratios.

\citet{Rigliaco2015ApJ...801...31R} reported the line ratio of H~\textsc{i} (9-7/)(7–6) ranging from 0.4 to 1.1, which are consistent with the models of \citetalias{Kwan2011MNRAS.411.2383K}. These ratios imply that H~\textsc{i} lines are likely probing gas with hydrogen number densities ($n_H$) between $10^{10}$ and $10^{11}$ cm$^{-3}$. In this study, we extend the analysis to MIR H~\textsc{i} lines using the \citetalias{Kwan2011MNRAS.411.2383K} models, following a similar approach to \citet{Antoniucci2017AA...599A.105A}. We adopt a 20-level hydrogen atom model with an assumed velocity gradient of $\mathrm{d}v/\mathrm{d}l = 150$~km~s$^{-1}$/2$R_\ast$ and an ionization rate of $\gamma_{\mathrm{H~\textsc{i}}} = 2 \times 10^{-4}$~s$^{-1}$ from \citet{Kwan2011MNRAS.411.2383K}. Figure~\ref{fig:7} presents the observed MIR line ratios from our sample, compared with the model predictions. The top panels show the model line ratios as a function of $n_{\rm H}$ for various gas temperatures, while the bottom panels display violin plots of the observed line ratios for each set of transitions.

Considering pairs of lines such as H~\textsc{i} (10--6)/H~\textsc{i} (7--6), the ratios rise steeply at $n_{\rm H} \sim 10^{10.4}$~cm$^{-3}$, marking the transition from optically thin to optically thick emission. {This optical depth effect makes line ratios such as H~\textsc{i} (10--6)/(7--6) highly sensitive tracers of high-density H~\textsc{i} gas.} This behaviour is also consistent with the result of \citet{Edwards2013ApJ...778..148E}, who attributed the turnover to collisional excitation and increasing optical depths at lower energy levels. Our analysis (Figure \ref{fig:7}) suggests that, for most young stars, $n_{\rm H}$ lies between $10^{10}$--$10^{10.5}$~cm$^{-3}$ for $T > 8750$~K, and between $10^{10.8}$--$10^{11.2}$~cm$^{-3}$ for $T \leq 8750$~K, broadly consistent with previous findings. However, estimating temperatures from these ratios is not possible due to degeneracies between temperature and density in determining the level emissivities. From Figure \ref{fig:7}, we can see that for $T = 8750$~K, the range of observed (shaded red region) $n_{\rm H}$ roughly matches for different sets of line ratios. Hence, we adopt $T = 8750$~K for the remainder of the analysis.

{The unprecedented sensitivity of JWST/MIRI allows for the simultaneous observation of multiple H~\textsc{i} transitions across different series, particularly higher-order (N$_{up} > 10$) MIR lines. This capability improves the accuracy of $n_{\rm H}$ determination, reducing degeneracies present in optical and NIR analyses}. {We selected the sources with at least 8 H~\textsc{i} line detection to perform this MIR decrement analysis.} Figure~\ref{fig:8} shows the H~\textsc{i} line ratios (normalized to H~\textsc{i}  6--5) as a function of hydrogen density, assuming a fixed temperature of $T = 8750$~K. The observed line ratios generally agree with the \citetalias{Kwan2011MNRAS.411.2383K} models for hydrogen densities in the range $n_{\rm H} = 10^{10.6}$--$10^{11.2}$~cm$^{-3}$, indicating predominantly optically thick emission. However, for a subset of sources, the observed ratios suggest emission from an optically thin regime with $n_{\rm H} < 10^{10}$~cm$^{-3}$.

{This analysis demonstrates that MIR H~\textsc{i} lines provide strong leverage for constraining the density of the emitting gas when compared with the optical and NIR series, using the \citetalias{Kwan2011MNRAS.411.2383K} model grids. In this work, we adopt $T=8750$~K as a representative value, but the degeneracy between $T$ and $n_{\rm H}$ remains evident: lower temperatures systematically produce higher inferred densities, whereas higher temperatures yield lower $n_{\rm H}$ for the same set of ratios. A natural extension of this approach in the future is a multi-wavelength analysis that incorporates optical, NIR, and MIR H~\textsc{i} lines, which would help to simultaneously constrain both temperature and density within the \citetalias{Kwan2011MNRAS.411.2383K} framework.}

\citet{Antoniucci2017AA...599A.105A} reported a correlation between the shape of the Balmer decrement and accretion rate, proposing that L-shaped (Type~4) decrements arise in high-accretion sources with $n_{\rm H} > 10^{11}$~cm$^{-3}$, while straight (Type~2) decrements occur in low-accretion sources with $n_{\rm H} \sim 10^{9}$~cm$^{-3}$. In contrast, our analysis reveals no such trend in the MIR. Sources exhibiting optically thin MIR H~\textsc{i} emission are not exclusively weak accretors. For example, CI~Tau (K4) and DR~Tau (K5) display similar MIR H~\textsc{i} decrement shapes, despite their accretion rates differing by nearly an order of magnitude. This suggests that additional factors -- such as the assumed temperature, spectral type, or magnetic field strength -- may influence the density of the emitting region.

As can be seen from the coloured \citetalias{Kwan2011MNRAS.411.2383K} curves in Figure~\ref{fig:8}, the $N_{\mathrm{low}} = 6$ and $7$ series are highly sensitive to changes in density above $n_{\rm H} = 10^{10.2}$~cm$^{-3}$. For $n_{\rm H} < 10^{10.2}$~cm$^{-3}$, the line ratios follow the standard decrement pattern, with lower-order transitions such as H~\textsc{i} (7--6) being strongest, and higher-order transitions (e.g. H~\textsc{i} (8--6), (9--6), (10--6)) following Case~B predictions. However, above this critical density, the lowest transition in the series (e.g. H~\textsc{i} (7--6) or (8--7)) is no longer dominant. This inversion becomes increasingly pronounced at higher densities ($n_{\rm H} > 10^{10.8}$~cm$^{-3}$). 

As seen in Figure \ref{fig:8}, typical gas densities in T~Tauri stars lie in the range $10^{10}$--$10^{11}$~cm$^{-3}$, H~\textsc{i} (7--6) becomes progressively weaker. At $n_{\rm H} = 10^{11}$~cm$^{-3}$, H~\textsc{i} (7--6) is nearly ten times weaker than H~\textsc{i} (10--6), indicating that higher-order lines of the Humphreys series are {enhanced} at higher densities. Similar behaviour is observed for transitions in the $N_{\rm low} = 7$ and $8$ series. This also explains the lower detection rate of H~\textsc{i} (8--7), even though it is the lowest transition in its series. At the densities observed for T~Tauri stars, the first transition in a given series is not necessarily the strongest. This effect is more pronounced in the MIR H~\textsc{i} ($N_{\rm low} = 6, 7,\dots$) series compared to the Balmer, Paschen, and Brackett series. {Complementing the MIR H~\textsc{i} lines with optical and NIR H~\textsc{i} transitions can help break the degeneracy between $T$ and $n_{\rm H}$ in the models of \citetalias{Kwan2011MNRAS.411.2383K}.}

\section{Summary}
\label{sec:summary}

We present a detailed analysis of mid-infrared (MIR) H~\textsc{i} emission lines in 79 Class~II protoplanetary discs using \textit{JWST}/MIRI spectroscopy. This study extends accretion diagnostics to higher-order hydrogen transitions, providing new insights into the inner disc regions with minimal contamination from jets and outflows.

The main results are summarized as follows:

\begin{itemize}

   \item A total of 22 MIR H~\textsc{i} transitions were homogeneously identified and measured across 79 Class~II protoplanetary discs. Among these, H~\textsc{i} (6--5) at 7.46~$\mu m$ was detected in 64 out of 79 sources, H~\textsc{i} (7--6) at 12.37~$\mu m$ in 61 sources, and H~\textsc{i} (8--6) at 7.50~$\mu m$ in 57 sources. Several higher-order transitions such as H~\textsc{i} (10--7) at 8.76~$\mu m$ (55/79) and H~\textsc{i} (9--6) at 11.31~$\mu m$ (42/79) were also robustly detected in a significant fraction of the sample, demonstrating the sensitivity of \textit{JWST}/MIRI to these previously unexplored accretion diagnostics.
    
    \item Spatial extent analyses using MIRI IFU observations confirm that H~\textsc{i} (6--5) and H~\textsc{i} (7--6) emissions originate from the inner accretion region within a radius of $\sim1\times$ PSF FWHM, clearly separating them from extended [Ne~\textsc{ii}] and [Fe~\textsc{ii}] jet emission.

    \item Molecular contamination, especially from H$_2$O, significantly affects several transitions. Median contamination levels reach $\sim75\%$ for H~\textsc{i} (12--7), $\sim60\%$ for H~\textsc{i} (7--6), and $\sim50\%$ for H~\textsc{i} (6--5). LTE slab modelling of H$_2$O was employed to correct the line fluxes and isolate accretion-related emission.

    \item {We provide updated empirical relations between MIR H~\textsc{i} line luminosities and accretion luminosities. Among the MIR H~\textsc{i} transitions analysed, the most reliable accretion diagnostics for T~Tauri stars with \textit{JWST}/MIRI are H~\textsc{i} (8--6) at 7.502~$\mu$m, H~\textsc{i} (10--7) at 8.760~$\mu$m, and H~\textsc{i} (9--6) at 5.908~$\mu$m. These lines are relatively free from strong molecular contamination and are less affected by optical depth effects, making them robust tracers of accretion in Class~II discs. In contrast, H~\textsc{i} (6--5) and (7--6), though intrinsically bright, are more susceptible to contamination from nearby H$_2$O features and therefore less reliable as stand-alone diagnostics.}

    \item {We show for the first time that the MIR H~\textsc{i} line ratios are the most appropriate tracer of high density H~\textsc{i} gas.} By comparing the observed MIR H~\textsc{i} line ratios to theoretical predictions from \citet{Kwan2011MNRAS.411.2383K}, the gas density in the emitting regions is constrained to $n_{\rm H} = 10^{10.6}$--$10^{11.2}$~cm$^{-3}$ for most sources. 
    
    \item We do not find evidence of a clear correlation between $n_{\rm H}$ and the accretion rate, which may reflect underlying variations in temperature, stellar properties, or magnetic field strength that are not captured in our current MIR analysis. {Complementary analyses using optical and NIR H~\textsc{i} lines will be essential to further constrain the physical conditions of the emitting gas.}

\end{itemize}

The high sensitivity and spectral resolution of \textit{JWST}/MIRI enable robust detections of multiple MIR H~\textsc{i} transitions and allow one to probe accretion processes in YSOs, particularly in embedded or distant systems inaccessible to optical and UV diagnostics. These results demonstrate the unique potential of mid-infrared spectroscopy with \textit{JWST} to advance our understanding of accretion physics across different stages of star formation. 

\section{Data availability}

The entirety of Tables \ref{tab:big_table} and \ref{tab:acc_estimates} are available in electronic form at the CDS via anonymous ftp to cdsarc.u-strasbg.fr (130.79.128.5) or via http://cdsweb.u-strasbg.fr/cgi-bin/qcat?J/A+A/.

\begin{acknowledgements}

{We thank the anonymous referee for their valuable suggestions, which have improved the flow of the manuscript}. This work is based on archival observations made with the NASA/ESA/CSA James Webb Space Telescope. The data were obtained from the Mikulski Archive for Space Telescopes at the Space Telescope Science Institute, which is operated by the Association of Universities for Research in Astronomy, Inc., under NASA contract NAS 5-03127 for JWST. These observations are associated with program ids: 1282, 1640,1676, 1549, 2025, 1584. The data described here can be obtained from \url{https://doi.org/10.17909/p4zy-g577}. B.S. acknowledges the TIFR postdoctoral fellowship. V.P., B.B., H.T. acknowledge the TIFR graduate fellowships. P.M. acknowledges the support of the Department of Atomic Energy, Government of India, under Project Identification No. RTI 4002. B.S. acknowledges the Infosys Leading Edge travel support. A.C.G. acknowledges support from PRIN-MUR 2022 20228JPA3A “The path to star and planet formation in the JWST era (PATH)” funded by NextGeneration EU and by INAF-GoG 2022 “NIR-dark Accretion Outbursts in Massive Young stellar objects (NAOMY)” and Large Grant INAF 2022 “YSOs Outflows, Disks and Accretion: towards a global framework for the evolution of planet forming systems (YODA)”.

\end{acknowledgements}

\bibliographystyle{aa} 
\bibliography{references} 

\begin{appendix}

\onecolumn

\section{Tables of stellar parameters from literature, newly measured accretion luminosity and accretion rates for the 79 sources studied in this work}

Table \ref{tab:big_table} shows the stellar parameters of the 79 sources used in this work. Distances are from \textit{Gaia DR3}, or from \citet{Manara2023ASPC..534..539M} and/or other literature for sources lacking \textit{Gaia DR3} distances. References for A$_V$, log(L$_{acc}$) and $\dot{M}_{\mathrm{acc}}\ (M_{\odot})\\$ are given in the table. Table \ref{tab:acc_estimates} shows the JWST-MIRI log($L_{\rm acc}$/$L_{\odot}$) and $\dot{M}_{\rm acc}$ values calculated using the empirical relations established in this work for four H~\textsc{i} transitions: (6--5), (7--6), (8--6), and (10--7). The line luminosity values are corrected for molecular contamination wherever possible.

\begin{table}[h!]
\tiny
\setlength{\tabcolsep}{3pt} 
\caption{Stellar parameters of the 79 stars studied in this work.} 
\label{tab:big_table}
\centering
\begin{tabular}{lccccccccccc}
\hline\hline
Source & RA & Dec & Sp. & Prop. & Dist. & $A_V$ & $\log \dot{M}_{\rm acc}$ & $L_\star$ & $\log L_{\rm acc}$ & Refs \\
Name & (deg) & (deg) & Type & ID & (pc) & (mag) & ($M_\odot\,\mathrm{yr}^{-1}$) & ($L_\odot$) & ($L_\odot$) & \\
\hline
*49 Cet & 23.658111 & -15.67636 & A1V & 1282 & $58.26^{+0.21}_{-0.23}$ & 0.11 & -7.8 & 14.8 & -0.26 & 21, 8 \\
CX Tau & 63.699442 & 26.80294 & M1.5Ve & 1282 & $126.73$ & 0.8 & ... & 0.34 & ... & 2 \\
CY Tau & 64.390555 & 28.34618 & M1.5 & 1282 & $124.35^{+0.79}_{-5.77}$ & 0.9 & -8.204 & 0.356 & -1.477 & 2, 1 \\
BP Tau & 64.815995 & 29.10733 & K5/7Ve & 1282 & $128.28^{+0.67}_{-0.60}$ & 0.5 & -7.29 & 0.978 & -0.559 & 2, 1 \\
FT Tau & 65.913297 & 24.93716 & M2.8 & 1282 & $129.96^{+0.35}_{-0.38}$ & 3.8 & -7.43 & 0.452 & 0.17 & 2, 17 \\
DF Tau & 66.761621 & 25.70608 & M3Ve & 1282 & $176.45$ & 0.1 & -7.55 & 0.35 & -0.75 & 3, 16 \\
DG Tau & 66.769548 & 26.10434 & K6Ve & 1282 & $130.21^{+1.86}_{-1.97}$ & 1.6 & -6.839 & 3.201 & -0.154 & 3, 1 \\
IQ Tau & 67.464804 & 26.11230 & M0.5 & 1640 & $130.72^{+0.77}_{-0.76}$ & 0.85 & ... & ... & ... & 3 \\
UX Tau & 67.516775 & 18.23030 & K2Ve+M1Ve & 1676 & $142.24$ & 0.7 & -7.96 & 2.2 & -0.55 & 6, 12 \\
V1213 Tau & 67.906160 & 18.20658 & K7 & 1282 & $140.00$ & 0.1 & ... & ... & ... & 27 \\

\hline
\end{tabular}
\tablebib{
(1)~\citet{Manara2023ASPC..534..539M}; (2)~\citet{Rigliaco2015ApJ...801...31R}; (3)~\citet{herczeg2014ApJ...786...97H}; (4)~\citet{Salyk2013ApJ...769...21S}; (5)~\citet{Evans2009ApJS..181..321E}; (6)~\citet{Furlan2009ApJ...703.1964F}; (7)~\citet{Mulders2017ApJ...847...31M}; (8)~\citet{Wichittanakom2020MNRAS.493..234W}; (9)~\citet{Manara2014AA...568A..18M}; (10)~\citet{Manara2015AA...579A..66M}; (11)~\citet{Dunham2015ApJS..220...11D}; (12)~\citet{Kim2013ApJ...769..149K}; (13)~\citet{Manara2017AA...604A.127M}; (14)~\citet{Manoj2006ApJ...653..657M}; (15)~\citet{Michel2021ApJ...921...72M}; (16)~\citet{Gangi2022AA...667A.124G}; (17)~\citet{Garufi2014AA...567A.141G}; (18)~\citet{Ribas2017ApJ...849...63R}; (19)~\citet{McDonald2017MNRAS.471..770M}; (20)~\citet{Salyk2013ApJ...769...21S}; (21)~\citet{Gontcharov2018MNRAS.475.1121G}; (22)~\citet{Sullivan2022AJ....164..100S}; (23)~\citet{Wahhaj2010ApJ...724..835W}; (24)~\citet{Yang2012ApJ...744..121Y}; (25)~\citet{Alcala2017AA...600A..20A}; (26)~\citet{Vioque2018AA...620A.128V}; (27)~\citet{Paunzen2024AA...689A.270P}; (28)~\citet{Hendler2017ApJ...841..116H}; (29)~\citet{ManzoMartinez2020ApJ...893...56M}; (30)~\citet{Yu2023ApJS..264...41Y}; (31)~\citet{Fernandes2023AJ....166..175F}; (32)~\citet{Luhman2020AJ....160...44L}; (33)~\citet{Dahm2009AJ....137.4024D}; (34)~\citet{Natta2006AA...452..245N}; (35)~\citet{Jayasinghe2018MNRAS.477.3145J}; (36)~\citet{Fairlamb2015MNRAS.453..976F}.
}
\end{table}

\begin{table}[h!]
\tiny
\setlength{\tabcolsep}{2pt} 

\caption{{Measured H~\textsc{i} line luminosities,} newly derived accretion luminosities and mass accretion rates.} 
\label{tab:acc_estimates}
\centering
\begin{tabular}{lcccc|cccc|cccc}
\hline\hline
\multirow{2}{*}{Target} & 
\multicolumn{4}{c}{$\log_{10}(L_{\mathrm{line}}/L_{\odot})$} & 
\multicolumn{4}{c}{$\log_{10}(L_{\mathrm{acc}}/L_{\odot})$} &
\multicolumn{4}{c}{$\dot{M}_{\mathrm{acc}}\ (M_{\odot}\ \mathrm{yr}^{-1})$} \\
\cline{2-5} \cline{6-9} \cline{10-13}
 & H~\textsc{i} (6--5) & H~\textsc{i} (7--6) & H~\textsc{i} (8--6) & H~\textsc{i} (10--7) & 
   H~\textsc{i} (6--5) & H~\textsc{i} (7--6) & H~\textsc{i} (8--6) & H~\textsc{i} (10--7) & 
   H~\textsc{i} (6--5) & H~\textsc{i} (7--6) & H~\textsc{i} (8--6) & H~\textsc{i} (10--7) \\
\hline
*49 Cet & \textless{}-6.67 & \textless{}-7.13 & \textless{}-6.69 & \textless{}-6.78 & \textless{}-2.53 & \textless{}-2.56 & \textless{}-2.46 & \textless{}-2.02 & \textless{}-9.41 & \textless{}-9.45 & \textless{}-9.35 & \textless{}-8.91 \\
CX Tau & -5.57(0.03) & -6.07(0.04) & -5.81(0.03) & -6.17(0.14) & -1.33(0.12) & -1.43(0.14) & -1.49(0.11) & -1.41(0.19) & -8.23(0.12) & -8.33(0.14) & -8.38(0.11) & -8.31(0.19) \\
CY Tau & -5.97(0.09) & -6.05(0.03) & -6.00(0.03) & \textless{}-6.42 & -1.76(0.18) & -1.41(0.14) & -1.71(0.13) & \textless{}-1.66 & -8.66(0.18) & -8.30(0.14) & -8.60(0.13) & \textless{}-8.55 \\
BP Tau & -4.87(0.02) & -5.42(0.07) & -5.22(0.02) & -5.58(0.08) & -0.57(0.09) & -0.74(0.11) & -0.85(0.08) & -0.83(0.13) & -7.48(0.09) & -7.65(0.11) & -7.76(0.08) & -7.74(0.12) \\
FT Tau & -4.96(0.04) & -5.50(0.06) & -5.10(0.01) & -5.64(0.02) & -0.68(0.09) & -0.83(0.11) & -0.71(0.08) & -0.88(0.10) & -7.56(0.09) & -7.71(0.11) & -7.59(0.08) & -7.76(0.10) \\
DF Tau & -4.76(0.23) & -4.91(0.11) & -4.66(0.06) & -5.06(0.02) & -0.46(0.26) & -0.20(0.15) & -0.23(0.12) & -0.31(0.11) & -7.38(0.26) & -7.12(0.15) & -7.14(0.12) & -7.22(0.11) \\
DG Tau & -4.36(0.05) & -4.59(0.05) & -4.30(0.04) & -4.62(0.04) & -0.02(0.13) & 0.14(0.14) & 0.16(0.14) & 0.12(0.15) & -6.94(0.13) & -6.77(0.14) & -6.75(0.14) & -6.79(0.15) \\
IQ Tau & -5.17(0.00) & -5.54(0.08) & -5.42(0.04) & -5.81(0.09) & -0.90(0.09) & -0.87(0.12) & -1.06(0.10) & -1.05(0.13) & -7.80(0.09) & -7.77(0.12) & -7.96(0.10) & -7.95(0.13) \\
UX Tau & -5.90(0.03) & \textless{}-6.31 & \textless{}-6.09 & \textless{}-6.27 & -1.69(0.15) & \textless{}-1.69 & \textless{}-1.80 & \textless{}-1.51 & -8.61(0.15) & \textless{}-8.61 & \textless{}-8.73 & \textless{}-8.43 \\
V1213 Tau & -6.26(0.08) & \textless{}-6.72 & \textless{}-6.47 & \textless{}-6.61 & -2.09(0.22) & \textless{}-2.13 & \textless{}-2.22 & \textless{}-1.85 & -9.00(0.21) & \textless{}-9.04 & \textless{}-9.13 & \textless{}-8.76 \\

\hline
\end{tabular}
\end{table}

\section{Continuum subtracted linemaps of H~\textsc{i}, H\textsubscript{2} and fine-structure lines of HV Tau C and RW Aur}

\begin{figure*}[h!]
    \centering
    \includegraphics[width=\columnwidth]{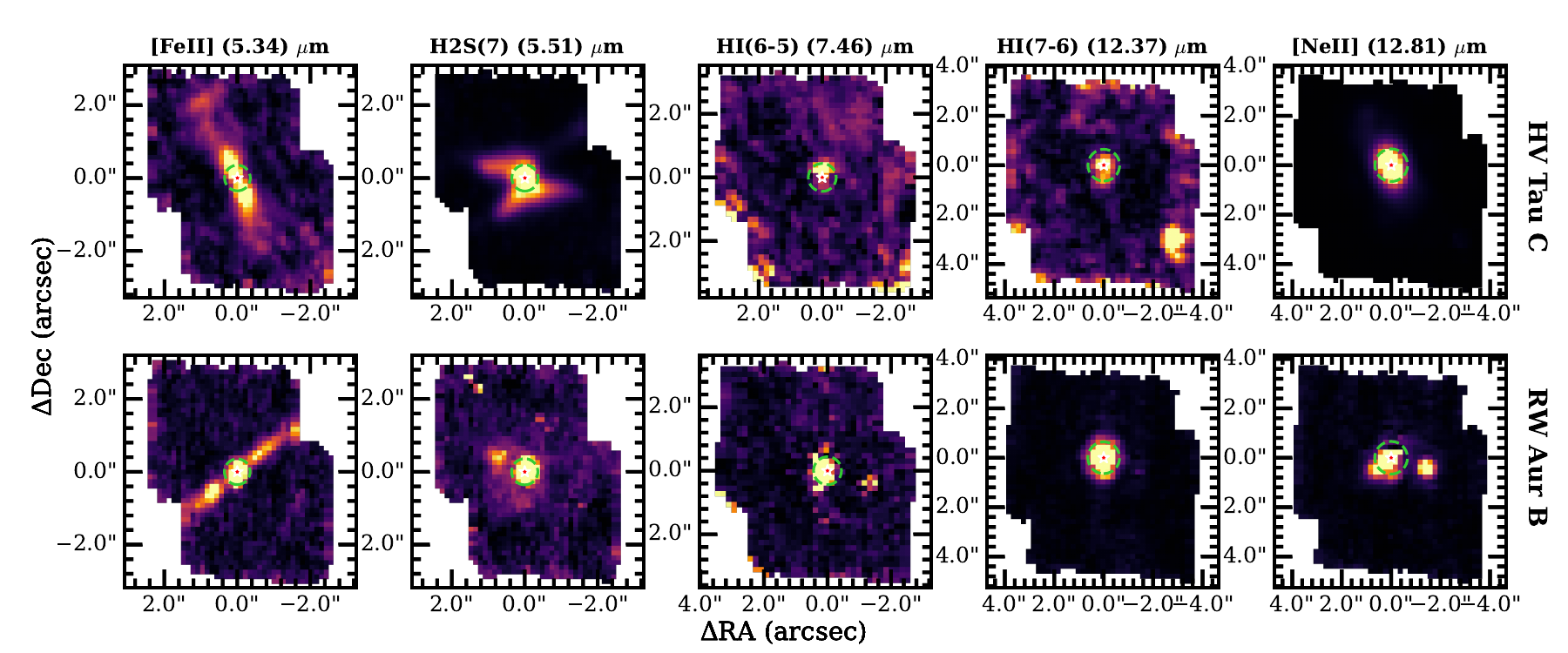}
    \caption{Continuum-subtracted emission-line maps. Respective maps of [Fe~\textsc{ii}], H$_2$~S(7), and selected H~\textsc{i} transitions for two representative sources: HV~Tau~C (top row) and RW~Aur~B (bottom row) are shown. The green circle marks the $2\times$PSF FWHM extraction region used for spectral analysis. The red star indicates the 14\,$\mu m$ continuum peak position. Despite the presence of extended outflows in [Fe~\textsc{ii}], the H~\textsc{i} lines are spatially unresolved within the $2\times$PSF region. Minor spatial shifts between cubes are due to residual pointing jitter (0.2$'$--0.3$'$) across spectral channels.}
    \label{fig:contsub_linemap}
\end{figure*}

Figure \ref{fig:contsub_linemap} shows the continuum-subtracted integrated emission-line maps of various lines for HV~Tau~C (top row) and RW~Aur~B (bottom row).  Both sources exhibit strong, collimated [Fe~\textsc{ii}] jets, while the H~\textsc{i} emission remains compact.

\FloatBarrier

\section{Impact of accretion variability and self-consistency of $\log(L_{\rm acc})$ measurements across different H~\textsc{i} lines}

\begin{figure*}
    \centering
    \includegraphics[width=\columnwidth]{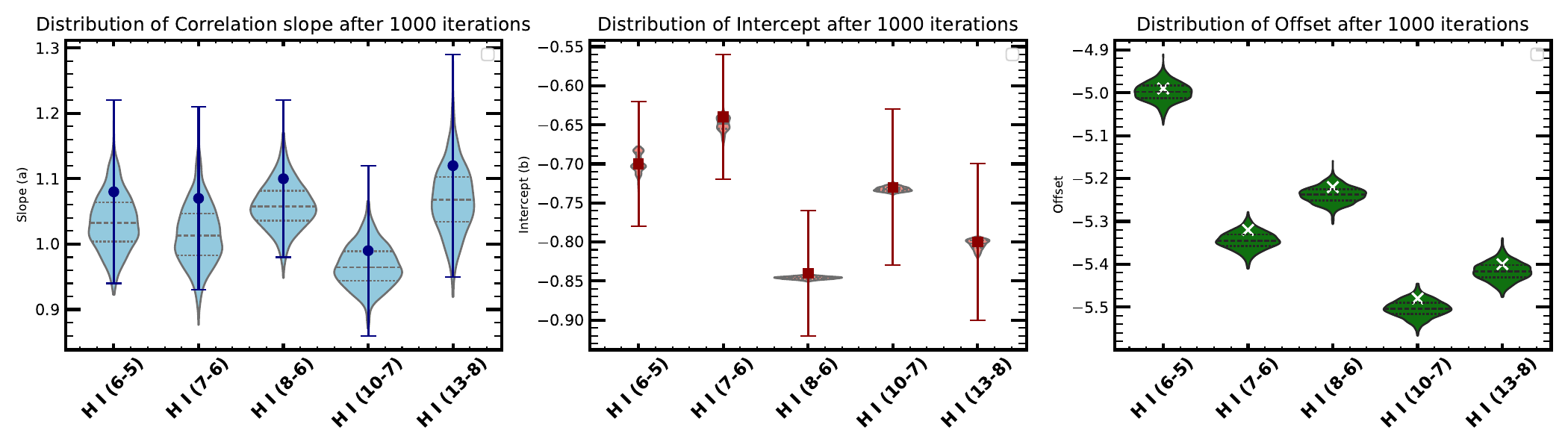}
    \caption{Effect of accretion variability. The sub-panels display the distributions of the correlation slope, intercept, and offset as violin plots from the Monte Carlo simulations, used to assess the impact of variability on the best-fit empirical relations in Table~\ref{tab:2}. The bars and 'x'-markers overlaid on the violin plots indicate the reported slope, intercept, and offset values for each line from Table~\ref{tab:2}. {The reported uncertainties incorporate the simulated $\pm50\%$ variability to the measured line fluxes to account for non-contemporaneous log($L_{\rm acc}$) measurements.}}
    \label{fig:monte_carlo_dist}  
\end{figure*}

{Figure~\ref{fig:monte_carlo_dist} shows the distributions of slope, intercept, and offset values for the different lines after 1000 Monte Carlo iterations, where the observed fluxes were varied by $\pm50\%$. This level of variability is meant to account for the uncertainty introduced by the non-contemporaneous $\log(L_{\rm acc})$ values adopted from the literature.}

\begin{figure*}[h!]
    \centering
    \includegraphics[width=\columnwidth]{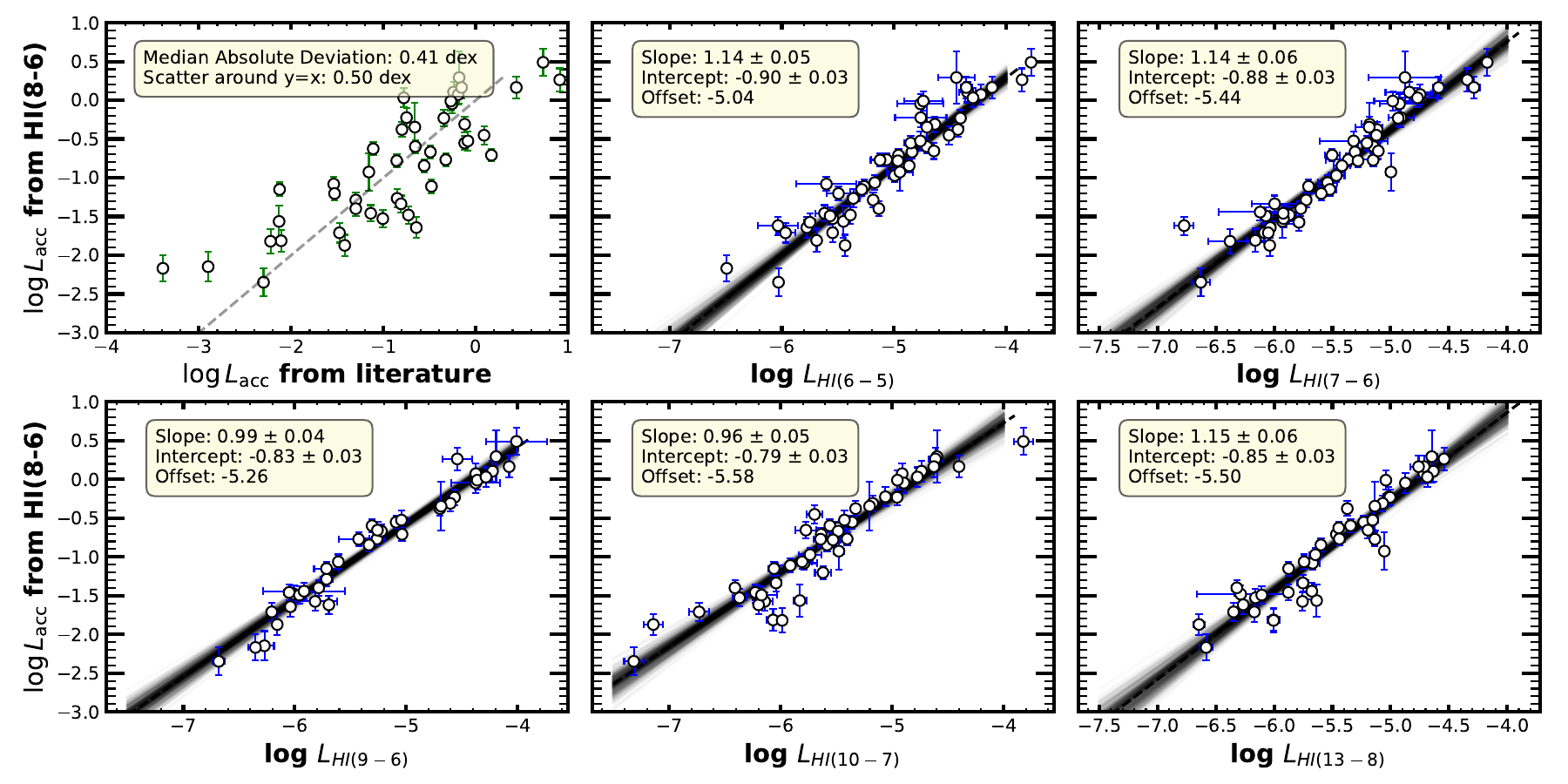}
    \caption{Possible effect of contemporaneous measurements. This plot shows the best-fit empirical relations for the H~\textsc{i} lines, recalibrated under the assumption that log($L_{\rm acc}$) derived from H~\textsc{i} (8--6) accurately provides the accretion, thereby mitigating the effects of non-simultaneous measurements. The resulting uncertainties on the empirical relations are smaller compared to those obtained using literature-based, non-contemporaneous $L_{\rm acc}$ values.}
    \label{fig:lacc_8-6}
\end{figure*}

\noindent Figure \ref{fig:lacc_8-6} shows the best-fit empirical relations of different H~\textsc{i} lines, assuming that the log($L_{\rm acc}$) derived from H~\textsc{i} (8--6) is a true estimate of accretion {at JWST's epoch}. The different sub-panels show that the observed scatter in Figure \ref{fig:5} has decreased significantly while using the log($L_{\rm acc~8-6}$).

\begin{figure*}
    \centering
    \includegraphics[width=\columnwidth]{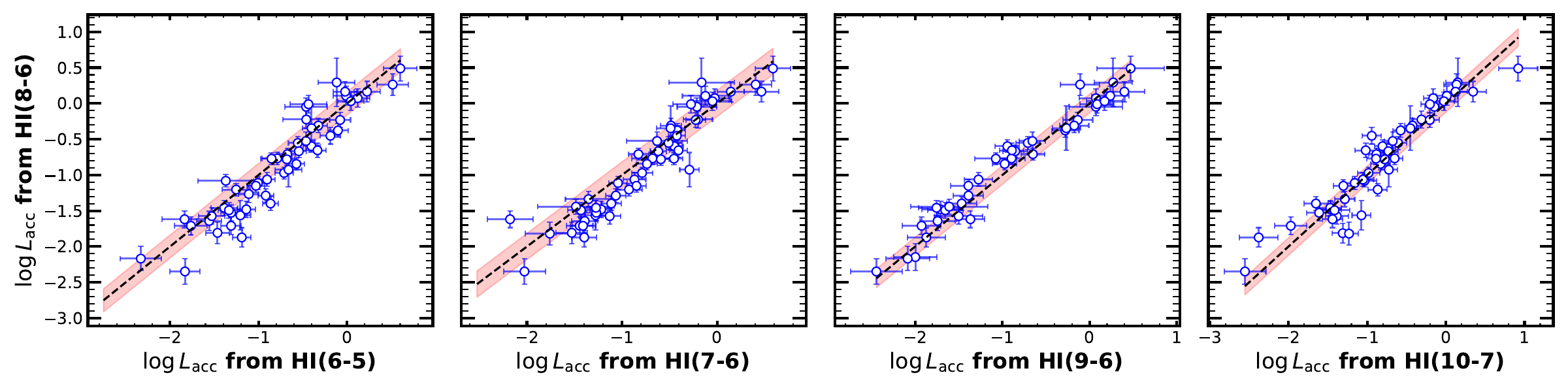}
    \caption{{Self-consistency of log($L_{\rm acc}$) measurements. The plot illustrates the consistency of the log($L_{\rm acc}$) values obtained from different H~\textsc{i} lines in our analysis (Table \ref{tab:acc_estimates}). The black dashed line represents the $y=x$ line, and the red shaded region highlights the median deviation (approximately 0.15 dex) from the $y=x$ line.}}
    \label{fig:lacc_cons}
\end{figure*}

{\noindent Figure~\ref{fig:lacc_cons} presents the scatter plot of log($L_{\rm acc}$) values derived from various H~\textsc{i} lines in our analysis (as detailed in Table~\ref{tab:acc_estimates}). The red-shaded region represents the median deviation (approximately 0.15 dex) from the $y=x$ dashed black line, indicating that the log($L_{\rm acc}$) measurements derived from different lines are consistent with each other, with a median deviation of 0.15 dex.}

\end{appendix}
\end{document}